%% file: uv_regions_v6.5.tex
\documentclass[iop, tighten]{emulateapj}

\pdfoutput=1  % pdflatex on arXiv

\usepackage{verbatim}  % comment env.
\usepackage{amsmath}  % substack

\defcitealias{Kang09}{K09}

\newlength{\dy}
\begin{document}

\title{The Panchromatic Hubble Andromeda Treasury. \textsc{VI}. The reliability
    of far-ultraviolet flux as a star formation tracer on sub-kpc scales.
}

\author{Jacob E. Simones\altaffilmark{1},
    Daniel R. Weisz\altaffilmark{2,3,8},
    Evan D. Skillman\altaffilmark{1},
    Eric F. Bell\altaffilmark{4},
    Luciana Bianchi\altaffilmark{5},
    Julianne J. Dalcanton\altaffilmark{3},
    Andrew E. Dolphin\altaffilmark{6},
    Benjamin D. Johnson\altaffilmark{2,7},
    Benjamin F. Williams\altaffilmark{3}
}

\altaffiltext{1}{Minnesota Institute for Astrophysics, University of Minnesota,
    116 Church Street SE, Minneapolis, MN 55455, USA; jsimones@astro.umn.edu;
    skillman@astro.umn.edu
}
\altaffiltext{2}{Department of Astronomy, University of California at Santa
    Cruz, 1156 High Street, Santa Cruz, CA 95064, USA; drw@ucsc.edu,
    bjohnso6@ucsc.edu
}
\altaffiltext{3}{Department of Astronomy, Box 351580, University of Washington,
    Seattle, WA 98195, USA; jd@astro.washington.edu; ben@astro.washington.edu
}
\altaffiltext{4}{Department of Astronomy, University of Michigan, 500 Church
    Street, Ann Arbor, MI 48109, USA; ericbell@umich.edu
}
\altaffiltext{5}{Department of Physics and Astronomy, Johns Hopkins University,
    3400 North Charles Street, Baltimore, MD 21218, USA; bianchi@pha.jhu.edu
}
\altaffiltext{6}{Raytheon, 1151 E. Hermans Road, Tucson, AZ 85756, USA;
    adolphin@raytheon.com
}
\altaffiltext{7}{Institut d'Astrophysique de Paris, 98 bis Boulevard Arago,
    75014 Paris, France
}
\altaffiltext{8}{Hubble Fellow}

\shortauthors{Simones et al.}

\begin{abstract}

We have used optical observations of resolved stars from the Panchromatic
Hubble Andromeda Treasury (PHAT) to measure the recent ($< 500\,\mathrm{Myr}$)
star formation histories (SFHs) of 33 FUV-bright regions in M31. The region
areas ranged from $\sim 10^4$ to $10^6\,\mathrm{pc}^2$, which allowed us to
test the reliability of FUV flux as a tracer of recent star formation on
sub-kpc scales. The star formation rates (SFRs) derived from the
extinction-corrected observed FUV fluxes were, on average, consistent with the
100-Myr mean SFRs of the SFHs to within the $1-\sigma$ scatter. Overall, the
scatter was larger than the uncertainties in the SFRs and particularly evident
among the smallest regions. The scatter was consistent with an even combination
of discrete sampling of the initial mass function and high variability in the
SFHs. This result demonstrates the importance of satisfying both the full-IMF
and the constant-SFR assumptions for obtaining precise SFR estimates from FUV
flux. Assuming a robust FUV extinction correction, we estimate that a factor of
2.5 uncertainty can be expected in FUV-based SFRs for regions smaller than
$10^5\,\mathrm{pc}^2$, or a few hundred pc. We also examined ages and masses
derived from UV flux under the common assumption that the regions are simple
stellar populations (SSPs). The SFHs showed that most of the regions are not
SSPs, and the age and mass estimates were correspondingly discrepant from the
SFHs. For those regions with SSP-like SFHs, we found mean discrepancies of
$10\,\mathrm{Myr}$ in age and a factor of 3 to 4 in mass. It was not possible
to distinguish the SSP-like regions from the others based on integrated FUV
flux.

\end{abstract}

\thanks{Based on observations made with the NASA/ESA Hubble Space Telescope,
    obtained from the Data Archive at the Space Telescope Science Institute,
    which is operated by the Association of Universities for Research in
    Astronomy, Inc., under NASA contract NAS 5-26555.
}


\section{Introduction}

A common technique for estimating global star formation rates (SFRs) in
individual galaxies is to measure the total flux at wavelengths known to trace
recent star formation (SF), such as ultraviolet (UV) emission from
intermediate- and high-mass stars. After correcting for dust extinction, an
observed flux can be converted into a SFR using a suitable calibration, which
is typically a linear scaling of intrinsic luminosity derived from population
synthesis modeling. The modeling process requires a set of stellar evolution
models and a stellar initial mass function (IMF), as well as a characterization
of the star formation history (SFH; the evolution of SFR over time) and the
metallicity of the population. These quantities are often not well-constrained
for a given system and need to be assumed (see reviews by
\citealt{Kennicutt98}, \citealt{Kennicutt12}, and references therein).

A set of flux calibrations widely used in extragalactic studies were presented
by \citet[][see \citealp{Kennicutt12} for updates]{Kennicutt98}. These
calibrations are based on models of a generic population with solar
metallicity, a fully populated IMF, and a SFR that has been constant over the
lifetime of the tracer emission ($\sim 100\,\mathrm{Myr}$ for UV). The flux
calibrations are therefore applicable to any population that can be assumed to
approximate the generic population, such as spiral galaxies. In environments
with low total SF (i.e., low mass) or on subgalactic scales, however, the
assumptions of a fully populated IMF and a constant SFR start to become
tenuous. As a result, applying the flux calibrations in these situations can
lead to inaccurate SFR estimates.

For populations located within a few Mpc, it is possible to measure SFRs more
directly by fitting the color magnitude diagram (CMD) of the resolved stars to
obtain a SFH \citep{Dolphin02}. At its core, CMD fitting is a population
synthesis technique just like flux calibration (albeit much more complex) and
thus requires a set of stellar evolution models, an IMF, and an accounting of
dust. The primary advantage of CMD fitting over the flux calibration method for
obtaining SFRs, however, is the elimination of assumptions about the SFH and
metallicity. CMD-based SFHs thus provide a relative standard for testing the
accuracy of SFR estimates from commonly used flux calibrations, especially in
applications where the underlying full-IMF and constant-SFR assumptions are not
strictly satisfied. More generally, the SFHs can be used to test results from
any other flux-based method, such as ages and masses derived under the simple
stellar population (SSP) assumption.

With recent Hubble Space Telescope (HST) observations from the Panchromatic
Hubble Andromeda Treasury \citep[PHAT;][]{Dalcanton12}, we have measured the
recent SFHs ($< 500\,\mathrm{Myr}$) of 33 UV-bright regions in M31 and compared
them with SFRs derived from UV flux. We also compared the SFHs with ages and
masses derived from UV flux by treating the regions as SSPs. The UV-bright
regions were cataloged by \citet[][\citetalias{Kang09} hereafter]{Kang09} using
Galaxy Evolution Explorer (GALEX) far-UV (FUV) flux and have areas ranging from
$10^4$ to $10^6\,\mathrm{pc}^2$. This range of sizes allows us to test the
reliability of the full-IMF, constant-SFR, and SSP assumptions on sub-kpc
scales.

This paper is organized as follows. We describe our sample of UV-bright regions
and show their CMDs from the PHAT photometry in \S \ref{observations}. We
summarize the CMD-fitting process, describe our extinction model, and present
the resulting SFHs of the regions in \S \ref{sfhs}. \S \ref{fluxmod} describes
the modeling of UV magnitudes from the SFHs, and \S \ref{sfrs} describes the
total masses and the mean SFRs from the SFHs, as well as the SFRs based on UV
flux. In \S \ref{discussion}, we compare the UV flux-based SFRs, ages, and
masses with the results from the SFHs, discuss the applicability of the
full-IMF, constant-SFR, and SSP assumptions to our sample, and attempt to
quantify the uncertainties associated with using UV flux to estimate SFRs,
ages, and masses for sub-kpc UV-bright regions.

% Figure 1
\begin{figure}
\centering
\includegraphics[width=\columnwidth]{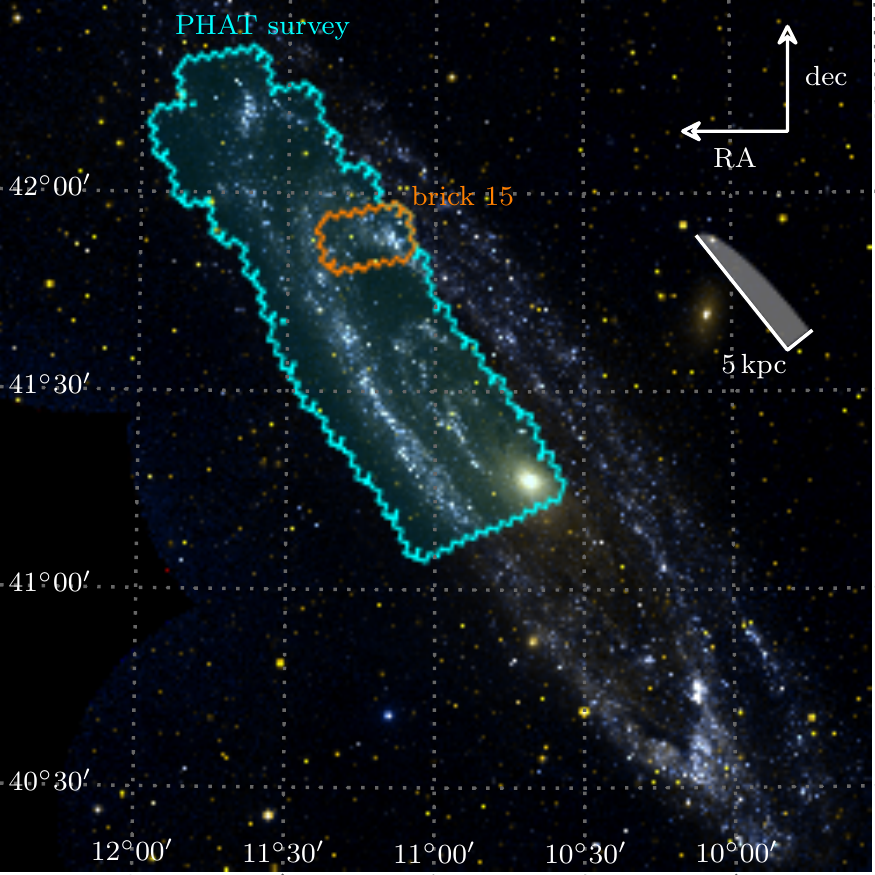}
\caption{Two-color composite mosaic of M31 from the GALEX Deep Imaging Survey
    (FUV in blue, NUV in orange). The HST/ACS outlines of the PHAT survey area
    and Brick 15 are highlighted in blue and orange, respectively. Brick 15
    covers a portion of the 10-kpc star-forming ring. The scale bar indicates a
    distance of $5\;\mathrm{kpc}$ along both the major and minor axes of M31
    assuming an inclination of $78\;\mathrm{deg}$ \citep{Tully94}.
}
\label{fig:map_full}
\end{figure}

% Figure 2
\begin{figure}
\centering
\includegraphics[width=\columnwidth]{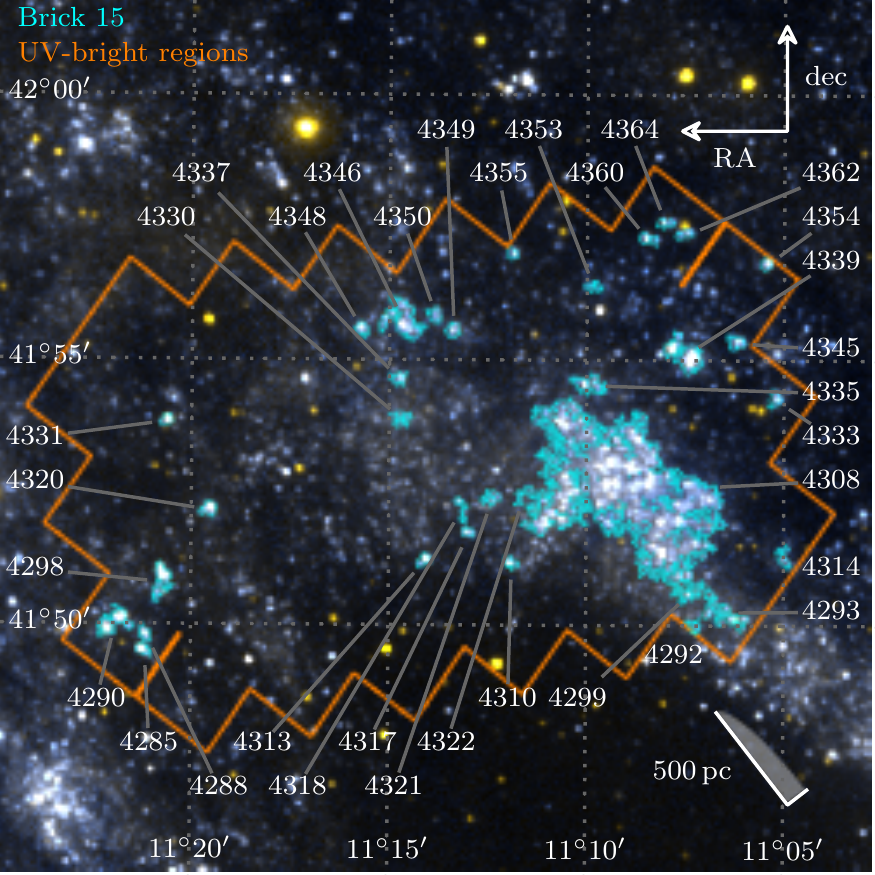}
\caption{Closeup of Brick 15 from the same image in Figure \ref{fig:map_full}.
    Brick 15 contains 33 of the UV-bright regions from the \citet{Kang09}
    catalog, highlighted in blue and labeled by ID number (see Table
    \ref{tab:observations}). The region areas, deprojected assuming an
    inclination of $78\;\mathrm{deg}$ \citep{Tully94}, range from $\sim 10^4$
    to $10^6\;\mathrm{pc}^2$. The scale bar indicates a distance of
    $500\;\mathrm{pc}$ along both the major and minor axes of M31.
}
\label{fig:map_b15}
\end{figure}

\section{Observations and photometry}\label{observations}

\subsection{UV-Bright Regions in M31}\label{observations.galex}

% First mention: Figure 1
% First mention: Figure 2
A set of UV-bright regions in M31 were defined by \citetalias{Kang09} using FUV
($\lambda \sim 1540\,\textrm{\AA}$) observations from GALEX. To summarize,
\citetalias{Kang09} defined a region as any area covering at least 50 contiguous
pixels ($113\,\mathrm{arcsec}^2$) with FUV surface brightness $\lesssim
25.9\,\mathrm{mag}\,\mathrm{arcsec}^{-2}$ (AB mag). For our sample, we selected
the subset of these regions that were within ``Brick 15'' of the PHAT survey, a
0.15-deg$^2$ area consisting of 18 individual fields, or HST pointings,
covering the 10-kpc star-forming ring (Figures \ref{fig:map_full} and
\ref{fig:map_b15}). Of all the bricks comprising the PHAT survey area, Brick 15
contains the greatest amount of SF and the largest number regions -- 33 total,
with respect to the combined outline of its Advanced Camera for Surveys (ACS)
images.

% First mention: Table 1
The identification numbers and locations of the regions in our sample as
reported in \citetalias{Kang09} are given in Table \ref{tab:observations}. For
each region, \citetalias{Kang09} measured the integrated FUV and NUV (near-UV,
$\lambda \sim 2320\,\textrm{\AA}$) magnitudes and subtracted the local
background estimated within a concentric annulus. We list the observed,
background-subtracted FUV magnitudes, $\mathrm{FUV_{obs}}$, and UV colors,
$\mathrm{(FUV-NUV)_{obs}}$ in Table \ref{tab:observations}. Table
\ref{tab:observations} also lists the solid angles and deprojected physical
areas of the regions, which we calculated assuming a distance to M31 of
$785\,\mathrm{kpc}$ \citep{McConnachie05} and a disk inclination of
$78\,\mathrm{deg}$ \citep{Tully94}. The areas range from $7.9 \times 10^3$ to
$7.3 \times 10^4\,\mathrm{pc}^2$, with one large outlier at $1.5 \times
10^6\,\mathrm{pc}^2$ (region 4308).

% Table 1
\input{table1}

\subsection{PHAT photometry}\label{observations.phat}

The resolved star photometry used in this study was taken from the PHAT Year 1
data release \citep{Dalcanton12}. The PHAT photometric catalogs were generated
using DOLPHOT, a version of HSTPHOT \citep{Dolphin00} with added ACS- and Wide
Field Camera 3-specific modules. Although the wavelength coverage of PHAT
extends from the UV to the near-infrared, we have used only the ACS optical
images (F475W and F814W filters) since they contain the greatest numbers of
stars and reach the deepest CMD features of the three PHAT cameras.

We applied quality cuts to the raw ACS photometric catalogs to minimize
non-stellar contaminants in our CMDs. Specifically, we required that each
object meet the following restrictions: $\mathtt{SNR}_\mathrm{F475W} \ge 4$,
$\mathtt{SNR}_\mathrm{F814W} \ge 4$, $(\mathtt{sharp}_\mathrm{F475W}^2 +
\mathtt{sharp}_\mathrm{F814W}^2) \le 0.075$, and
$(\mathtt{crowd}_\mathrm{F475W}^2 + \mathtt{crowd}_\mathrm{F814W}^2) \le 1.0$,
where $\mathtt{SNR}$, $\mathtt{sharp}$, and $\mathtt{crowd}$ refer to the
DOLPHOT signal-to-noise, sharpness, and crowding parameters in each filter.
These quality cuts are the ``gst'' cuts outlined in the main PHAT data release
\citep{Dalcanton12}.

\begin{figure*}
\centering
\includegraphics[width=\textwidth]{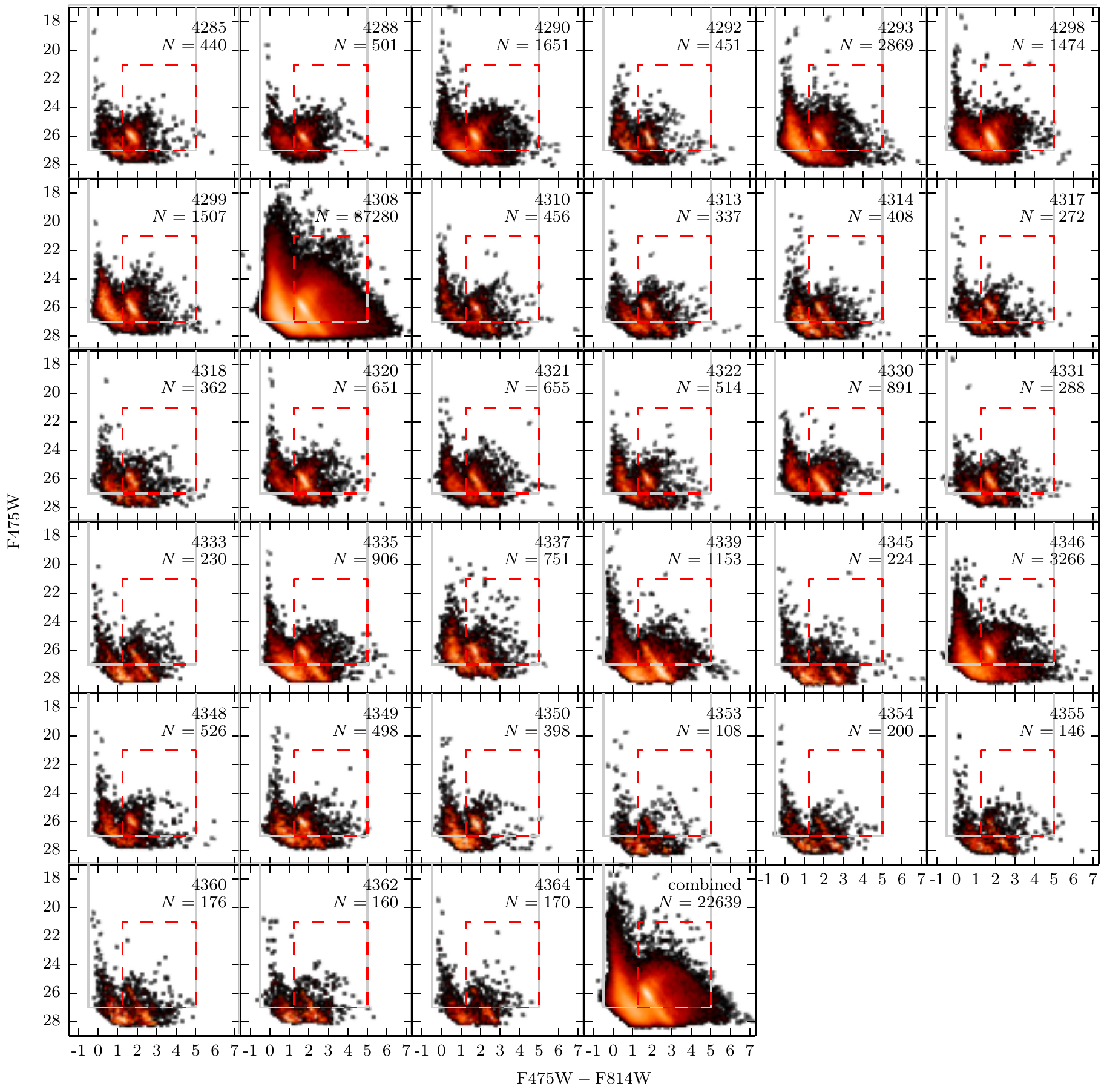}
\caption{Optical color-magnitude diagrams (ACS/WFC filters F475W and F814W) of
    the UV-bright regions. Region ID numbers and the numbers of stars fit by
    MATCH are shown in each panel. The MATCH fit area is inside the solid gray
    line, where the faint end marks the 50\% completeness limit in F475W. Stars
    within the red dashed line were excluded from the fit. The combined region
    includes all regions except 4308. The CMDs show broadening of the main
    sequence and other features, indicating that the regions are subject to
    nontrivial amounts of differential extinction from dust internal to M31.
}
\label{fig:cmd_grid}
\end{figure*}

% First mention: Figure 3
We extracted all stars within the boundaries of the 33 UV-bright regions,
combining photometry as needed for regions extending across multiple ACS
fields. We did not take advantage of the improved signal-to-noise ratio where
fields overlapped. The CMDs of the Brick 15 UV-bright regions are shown in
Figure \ref{fig:cmd_grid}.

% Figure 3

\subsection{Artificial star tests}\label{observations.asts}

To assess observational errors and characterize photometric completeness, we
conducted $\sim 2.5 \times 10^4$ artificial star tests (ASTs) for each of the
regions. The color and magnitude distributions for the ASTs were modeled after
the CMDs of the individual regions. However, as discussed below in \S
\ref{sfhs.extinction}, we excluded the red giant branch (RGB) and red clump
(RC) from the SFH analysis. We therefore only considered ASTs with properties
similar to the blue portion of the CMDs, including the luminous main sequence
(MS).

We used the ASTs to compute the photometric completeness functions for each of
the 33 regions. The completeness functions were consistent throughout the
sample, with an uncertainty of $0.06\,\mathrm{mag}$ in the mean 50\%
completeness limit in each filter. In addition, the photometric errors varied
little between the regions. These consistencies allowed us to combine the ASTs
from the individual regions for a total of $1.6 \times 10^6$ ASTs. This
hundredfold increase in the number of ASTs available to each region provided a
superior CMD error model for the SFH measurement process. The 50\% completeness
limits of the region sample are $27.0\,\mathrm{mag}$ in F475W and
$26.2\,\mathrm{mag}$ in F814W.

\section{The recent SFHs of UV-bright regions in M31}\label{sfhs}

The derivation of the SFHs for the UV-bright regions is described in this
section. The first subsection gives a brief discussion of the SFH code and
describes the overall SFH measurement procedure from beginning to end. Details
of the extinction model, the resulting SFHs, and our uncertainty analysis are
discussed in the subsequent subsections.

\subsection{CMD modeling with MATCH}\label{sfhs.measurement}

We used the SFH code MATCH \citep{Dolphin02} to measure the SFHs of our sample
of UV-bright regions. Assuming a stellar IMF, binary fraction, and a set of
stellar evolution models, MATCH constructs a series of synthetic CMDs over
given ranges in distance, age, metallicity, and extinction. The synthetic CMDs
are convolved with the error model from the ASTs to account for observational
errors. Linear combinations of the synthetic CMDs form a model which is
assigned a fit value based on a comparison with the observed CMD. The SFH of
the model CMD that minimizes the fit value is considered the most likely SFH of
the observed population given the input parameters. We emphasize that MATCH
models the \emph{distribution} of stars in the observed CMD, not the ages and
masses of the individual stars.

The fit statistic used by MATCH is equal to $-2 \ln \Lambda_P$, where
$\Lambda_P$ is the Poisson likelihood ratio. According to Wilks' theorem, this
statistic is asymptotically $\chi^2$ distributed, allowing us to estimate the
$n\sigma$ confidence limits in a set of SFH solutions using the condition
$\mathrm{fit-fit_{min}} \le n^2$, where $\mathrm{fit_{min}}$ corresponds to the
best-fit SFH. This method was used to estimate various uncertainties in \S
\ref{sfhs.results} and \S \ref{sfhs.uncertainties}.

We assumed the following for our SFH measurements:
\begin{enumerate}
\item A Kroupa IMF \citep{Kroupa01}.
\item The Padova stellar evolution models for masses between 0.15 and
    $120\,M_\odot$ (the IMF was normalized using masses down to
    $0.01\,M_\odot$) including updated low-mass asymptotic giant branch tracks
    \citep{Girardi10}.
\item A binary fraction of 0.35 with a uniform secondary mass distribution.
\item A distance modulus of 24.47 \citep{McConnachie05}. The distance to M31 is
    fairly well-known, allowing us to fix this value and eliminate a free
    parameter in the CMD fitting process.
\item A set of 48 log-spaced age bins from $\log_{10}(\mathrm{Age/yr}) = 6.60$
    to $9.00\,\mathrm{dex}$ with width $\Delta \log_{10}(\mathrm{Age/yr}) =
    0.05\,\mathrm{dex}$ (though as discussed in \S \ref{sfhs.results}, we
    ultimately only consider the SFH out to $500\,\mathrm{Myr}$, or
    $\log_{10}(\mathrm{Age/yr}) = 8.70$).
\item A metallicity range of $[\mathrm{M/H}] = -2.3$ to $0.1\,\mathrm{dex}$ at
    a resolution of $0.1\,\mathrm{dex}$ with the requirement of a monotonically
    increasing chemical evolution model.
\end{enumerate}

We also simulated the effects of intervening Galactic foreground populations
using the TRILEGAL population synthesis model \citep{Girardi05}. The solid
angles of the regions were small enough, however, that no more than a few
foreground stars were expected per CMD, implying a negligible impact on our
final results.

\begin{figure*}
\centering
\includegraphics[width=\textwidth]{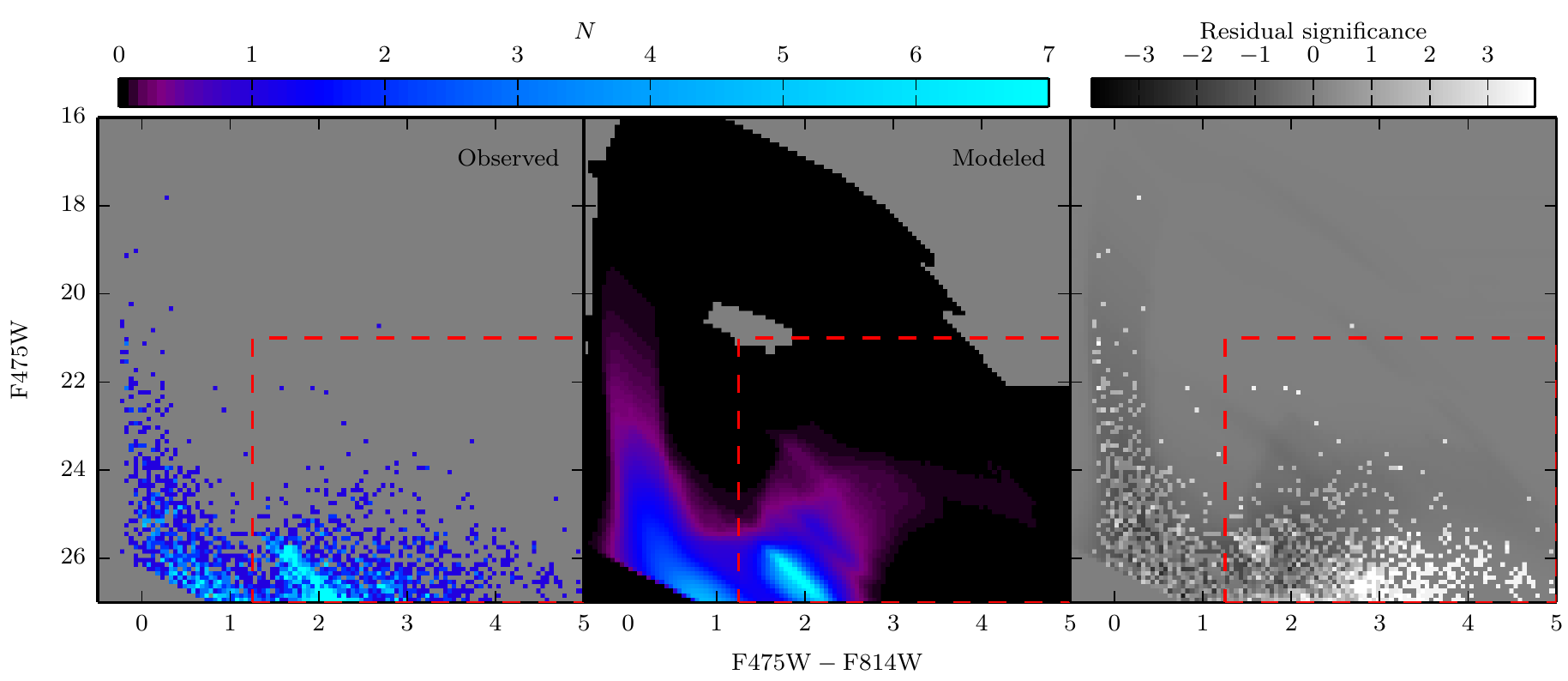}
\caption{Observed CMD of region 4339, with the best-fit modeled CMD and the
    residual significance (the observed CMD minus the modeled CMD, weighted by
    the variance). The color maps indicate the number of stars in each color
    and magnitude bin. The CMD limits correspond to the solid gray lines shown
    in Figure \ref{fig:cmd_grid}, and the dashed red line shows the area
    excluded from the fitting process. We find no systematic residuals in the
    MS, indicating that the model is a good fit to the data.
}
\label{fig:cmdfit_4339}
\end{figure*}

% First mention: Figure 4
Extinction was modeled using two parameters, $A_{\mathrm{V}f}$ and
$dA_\mathrm{V}$, as described in \S \ref{sfhs.extinction}. For each region, we
sampled the extinction parameter surface using a combination of pattern search
and grid search techniques, measuring the best-fit SFH at each point. The
search procedure resulted in an irregularly-sampled grid of MATCH fit values
with a minimum step size of $0.05\,\mathrm{mag}$ in both $A_{\mathrm{V}f}$ and
$dA_\mathrm{V}$.\footnote{Computing fully-sampled grids for the regions at
$0.05\,\mathrm{mag}$ resolution over reasonable ranges in $A_{\mathrm{V}f}$ and
$dA_\mathrm{V}$ was found to be computationally infeasible.} We then compared
the fit values across the grid to find the overall best-fit SFH. Figure
\ref{fig:cmdfit_4339} shows an example model CMD for the best-fit SFH of region
4339, along with the observed CMD and the residual significance.

% Figure 4

\subsection{Extinction model}\label{sfhs.extinction}

\citetalias{Kang09} measured the average $E(\mathrm{B-V})$ reddening in each
region using the reddening-free parameter $Q$ and UBV photometry for individual
OB stars, providing us with possible constraints on extinction for CMD fitting
with MATCH. However, the CMDs in Figure \ref{fig:cmd_grid} show broadening of
the intrinsically narrow MS, indicating that the regions are subject to
nontrivial amounts of differential extinction from dust internal to M31. In
some regions the differential extinction is severe enough that the MS appears
doubled. Differential extinction is also evident in the population of older
stars, which we assume to be reasonably well-mixed throughout the galaxy,
characterized by a broad RGB and an elongation of the RC along the reddening
vector. These complexities lead to poor results when fitting an entire CMD with
a single extinction value, such as that obtained from the average
$E(\mathrm{B-V})$ in a region.

To fit the CMDs more accurately, we adopted a two-parameter extinction model
consisting of a foreground dust component and a differential component. The
total V-band extinction common to all stars in the CMD is set by the foreground
parameter, $A_{\mathrm{V}f}$. Differential extinction is added to the stars in
varying amounts following a uniform distribution from zero up to a maximum
determined by the differential parameter, $dA_\mathrm{V}$. Compared to the
simplest case of optimizing a single extinction parameter, this extinction
model provided much better fits for the observed CMDs while allowing MATCH to
compute best-fit SFH solutions in a reasonable amount of time.

A specific shortcoming of the model, however, is that not all populations are
expected to have the same extinction profile. Young stars tend to reside closer
to the midplane of the galaxy and are likely to be physically associated with
cold dense gas that hosts the dusty ISM. The older RGB and RC stars, which
dominate the CMDs of the regions, can have a much larger scale height in
comparison. To prevent the older populations from influencing the parameters of
the extinction model we excluded all stars with both $\mathrm{F475W-F814W} >
1.25$ and $\mathrm{F475W} > 21.0\,\mathrm{mag}$ (red dashed lines in Figure
\ref{fig:cmd_grid}) from the CMD fitting process. The SFHs and extinction
parameters we derive from MATCH therefore correspond only to the distributions
of massive MS stars (the primary producers of UV flux) as well as any blue and
red He-burning stars in the CMDs.

By creating an exclusion area in the CMD, we necessarily place a limit on the
total extinction that can be determined by MATCH. From the CMDs in Figure
\ref{fig:cmd_grid}, the maximum amount of reddening a MS star can have before
entering the exclusion area is $\mathrm{F475W-F814W} \approx
1.7\,\mathrm{mag}$. Assuming the extinction curve from \citet[][see \S
\ref{discussion.fuv}]{Cardelli89}, this amount of reddening corresponds to a
total extinction of $A_{\mathrm{V}f} + dA_\mathrm{V} \approx
2.8\,\mathrm{mag}$. CMD models with total extinction at this limit are
indistinguishable from higher-extinction models because stars in the exclusion
area do not affect the MATCH fit statistic. We therefore place an upper limit
of $2.8\,\mathrm{mag}$ on the total extinction, $A_{\mathrm{V}f} +
dA_\mathrm{V}$, during the optimization of the SFHs described in \S
\ref{sfhs.measurement}.

One caveat for our two-component model is that observational studies of the ISM
routinely demonstrate log-normal, not uniform, density distributions
\citep[e.g.,][]{Berkhuijsen08, Hill08, Ballesteros11, Shetty11, Dalcanton14}.
Modeling the extinctions in M31 with log-normal distributions has been
successful for producing extinction maps that agree with the emission from dust
and gas \citep{Dalcanton14}. Implementing such a model in MATCH would require a
minimum of three parameters: a foreground component, and the mean and variance
for the log-normal. A more realistic extinction model might account for the
fraction of stars affected by the log-normal as well as the scale height of the
stars relative to the gas in the disk, which can vary with age. With each
additional parameter, however, the size of the search space increases
exponentially and measuring the SFH of a single region quickly becomes
impractical. It is difficult to assess how the derived SFHs are affected by our
comparatively simple extinction model without repeating the measurements with a
more sophisticated model. Even so, the quality of the residuals for the modeled
CMDs (e.g., Figure \ref{fig:cmdfit_4339}) suggests that the two-component model
is reasonably accurate.

\subsection{Results}\label{sfhs.results}

% First mention: Figure 5
% First mention: Table 2
We present the SFHs of the UV-bright regions in Figure \ref{fig:sfh_grid}. The
corresponding best-fit $A_{\mathrm{V}f}$ and $dA_\mathrm{V}$ parameters are
listed in Table \ref{tab:SFHdata}. The uncertainties of the parameters for each
region correspond to the minimum and maximum values among the set of SFHs
within $1\sigma$ of the best-fit SFH on the $A_{\mathrm{V}f},dA_\mathrm{V}$
surface (i.e., all SFHs for which $\mathrm{fit-fit_{min}} \le 1$; see \S
\ref{sfhs.measurement}). The final metallicities of the best-fit SFHs for all
regions ranged from $-1.30\,\mathrm{dex} \le [{\rm M/H}] \le
0.01\,\mathrm{dex}$, with 80\% of the values within $0.3\,\mathrm{dex}$ of the
mean, $[{\rm M/H}] = -0.3\,\mathrm{dex}$.

The exclusion area in the CMDs and the 50\% photometric completeness limit both
restrict the age of the oldest population that can be fit by MATCH. Through
synthetic CMD modeling, we find that a significant fraction of the stars in
populations older than $\sim 500\,\mathrm{Myr}$ are either within the exclusion
area or below the 50\% photometric completeness limit. In comparison, younger
populations are well-represented in the MS/He-burning area of the CMD. We
therefore adopt $500\,\mathrm{Myr}$ as the maximum reliable age of the SFHs.

Considering that the UV emission from an SSP becomes negligible after $\sim
100\,\mathrm{Myr}$ \citep{Gogarten09, Leroy12}, we chose to display only the
past $200\,\mathrm{Myr}$ of the SFHs. This was done to show as much of the
overall history as possible while preserving sufficient detail in the
$0-100\,\mathrm{Myr}$ range. Also, the SFHs are shown at a coarser time
resolution than the actual resolution of $\Delta \log_{10}(\mathrm{Age/yr}) =
0.05\,\mathrm{dex}$ to simplify visual comparisons between the regions.
\emph{We use the full-resolution SFHs for all analyses that follow.}

The Padova stellar evolution models used to fit the CMDs do not include ages
less than $4\,\mathrm{Myr}$, creating a gap between the present time and the
youngest age bin in the SFHs. To account for this, we extended the youngest bin
to cover the ages in the gap and rescaled its SFR such that the total mass
formed in the bin was conserved.

% Figure 5
\begin{figure*}
\centering
\includegraphics[width=\textwidth]{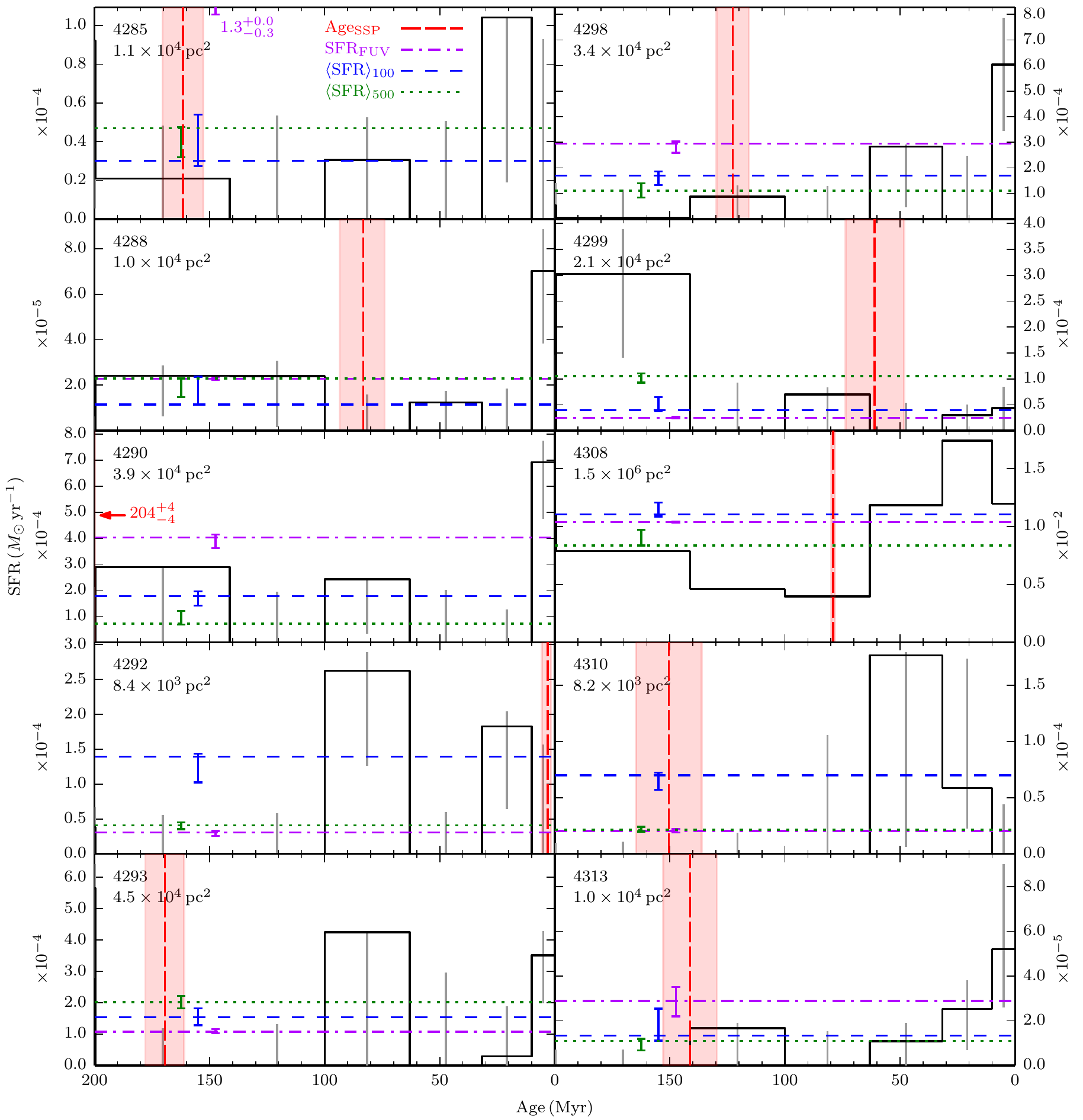}
\caption{SFHs of UV-bright regions in M31 (black histogram). The SFH for the
    combined region in Figure \ref{fig:cmd_grid} was derived independently. The
    region ID number and deprojected area are given in each panel. The vertical
    long-dashed red line shows $\mathrm{Age_{SSP}}$, the SSP age from
    \citet{Kang09}, which does not accurately describe the majority of the
    SFHs. The dashed-dotted purple line shows the constant SFR,
    $\mathrm{SFR_{FUV}}$, obtained from the extinction-corrected observed FUV
    fluxes. The short-dashed blue and dotted green lines show $\langle
    \mathrm{SFR}\rangle_{100}$ and $\langle \mathrm{SFR}\rangle_{500}$, the
    mean SFRs over the last 100 and $500\,\mathrm{Myr}$, respectively.
}
\label{fig:sfh_grid}
\end{figure*}

% Figure 5, continued: page 6
\begin{figure*}
\begin{center}
\figurenum{\ref{fig:sfh_grid}}
\includegraphics[width=\textwidth]{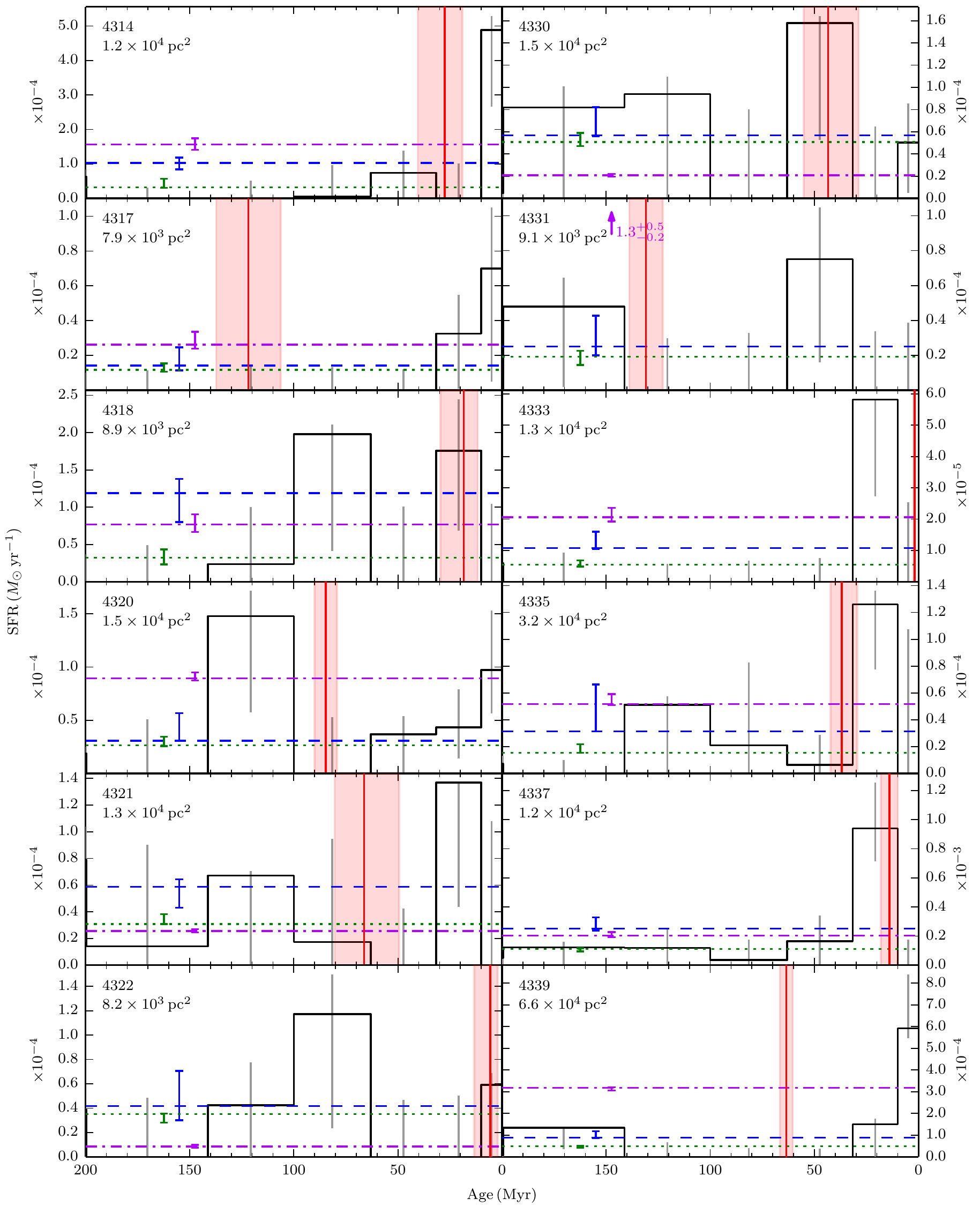}
\caption{\textit{Cont.}}
\end{center}
\end{figure*}

% Figure 5, continued: page 7
\begin{figure*}
\begin{center}
\figurenum{\ref{fig:sfh_grid}}
\includegraphics[width=\textwidth]{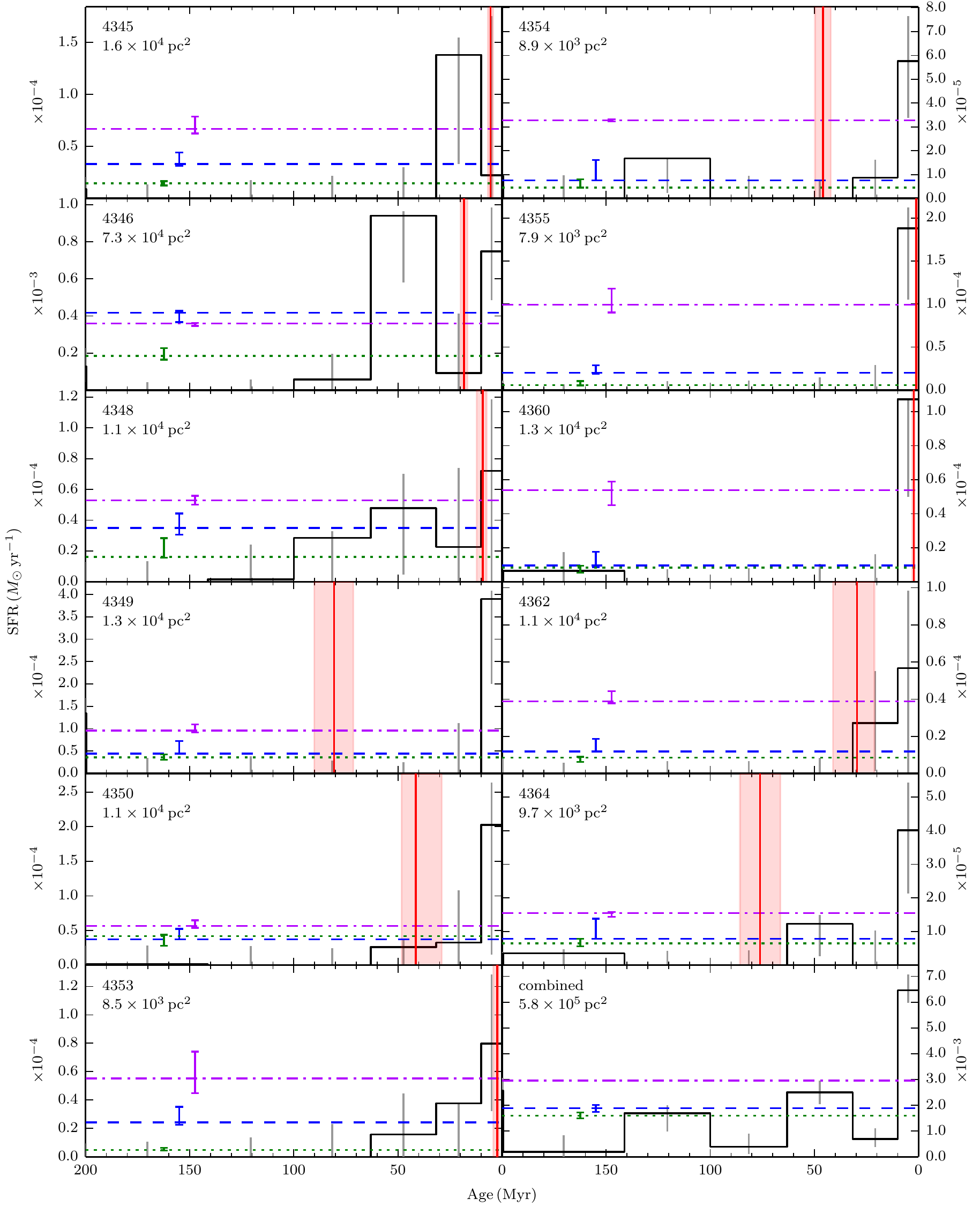}
\caption{\textit{Cont.}}
\end{center}
\end{figure*}

% Table 2
\input{table2}

\subsection{Uncertainties}\label{sfhs.uncertainties}

The random uncertainties of the SFHs were evaluated using the Hybrid Monte
Carlo (HMC) method described in \citet{Dolphin13}. With this method, the SFH
probability density function (PDF) is estimated from a sequence (or chain) of
samples in SFH space, which is parameterized by age, metallicity, and SFR. Each
sample in the chain is proposed and then either accepted or rejected using
Hamiltonian dynamics to efficiently obtain a set of samples that are
distributed according to the underlying PDF. For each region, we ran the HMC
algorithm for a total of $10^4$ accepted proposals and calculated the $1\sigma$
random uncertainties from the narrowest interval containing 68\% of the area
under the PDF.

The process of minimizing the SFHs with respect to the extinction model
resulted in irregular grids of fit values on the
$A_{\mathrm{V}f},dA_\mathrm{V}$ surface (\S \ref{sfhs.measurement}). For each
region, we selected all SFHs in the grid within $1\sigma$ of the best-fit SFH
using the condition $\mathrm{fit-fit_{min}} \le 1$. The distribution for this
set of SFHs was then used to estimate the $1\sigma$ systematic uncertainties in
the best-fit SFH related to the measurement of $A_{\mathrm{V}f}$ and
$dA_\mathrm{V}$.

The error bars in Figure \ref{fig:sfh_grid} correspond to the combination of
the random and systematic uncertainties. We did not assess the systematic
uncertainties related to the stellar evolution models used with MATCH.

We use the HMC tests and the ``$1\sigma$'' set of SFHs to estimate the random
and systematic uncertainties for all quantities derived from the SFHs (FUV
magnitudes, total masses, etc.). For example, the mass of recently-formed stars
in a region (see \S \ref{sfrs}) was calculated for all of the HMC SFHs, and the
random uncertainty was calculated from the distribution of the resulting
masses. The systematic uncertainty was estimated from the minimum and maximum
masses derived from the set of $1\sigma$ SFHs for the region. We then added the
random and systematic components in quadrature to get the total uncertainty for
the mass of the best-fit SFH.

\section{UV flux modeling}\label{fluxmod}

We used the SFHs in Figure \ref{fig:sfh_grid} as a basis for modeling the total
present-day UV fluxes for each region. This technique was pioneered by
\citet{Gogarten09} in their study of UV-bright regions in the outer disk of
M81, and has recently been extended to several dozen dwarf galaxies in the
Local Volume \citep{Johnson13}.

Following the procedure described in \citet{Johnson13}, the intrinsic
(unreddened) FUV and NUV fluxes were modeled from the SFHs using the Flexible
Stellar Population Synthesis (FSPS) code \citep{Conroy09, Conroy10}. FSPS was
run using the Padova isochrones \citep{Girardi10} and a Kroupa IMF
\citep{Kroupa01}. The metallicity for all regions was set to a constant
$[\mathrm{M/H}] = -0.3\,\mathrm{dex}$, based on the approximate final
metallicities of the SFH solutions (\S \ref{sfhs.results}). The effect of
assuming a homogeneous metallicity value is discussed in \S
\ref{discussion.fuv}.

% First mention: Table 3
The modeled FUV and NUV fluxes were converted into AB magnitudes using the
formulae in \citet{Morrissey07}, and the uncertainties were calculated as
described in \S \ref{sfhs.uncertainties}. The intrinsic FUV magnitudes,
$\mathrm{FUV_{SFH,0}}$, and UV colors, $\mathrm{(FUV-NUV)_{SFH,0}}$ of the
regions are listed in Table \ref{tab:fluxmod}.

\begin{figure}
\centering
\includegraphics[width=\columnwidth]{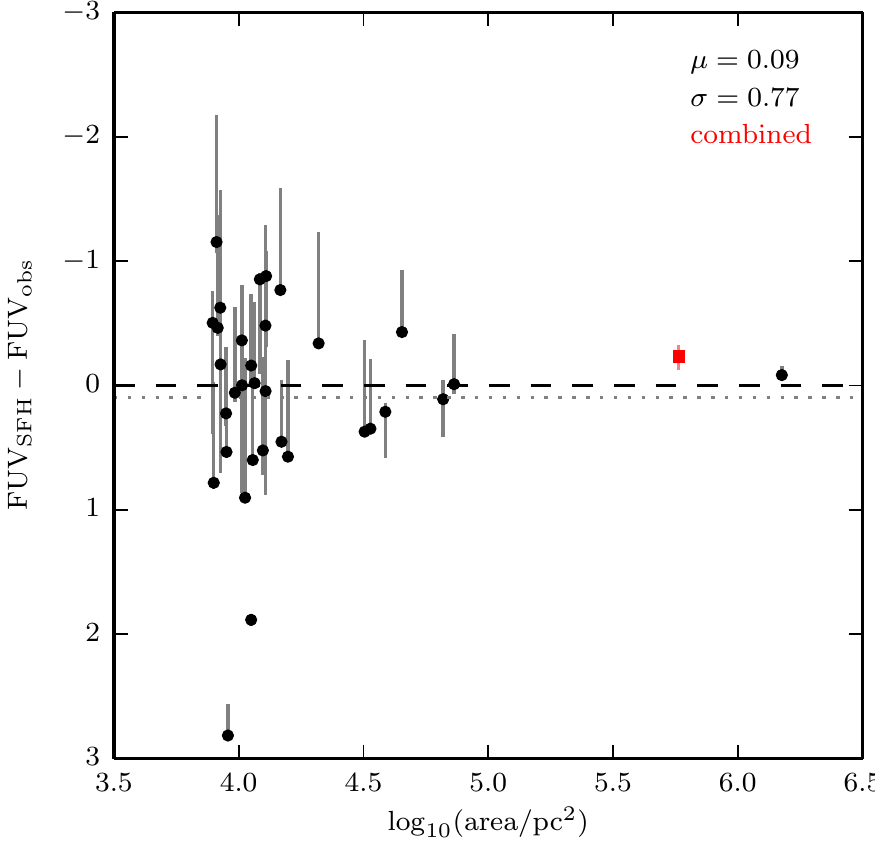}
\caption{The difference between the modeled reddened ($\mathrm{FUV_{SFH}}$) and
    observed \citep[$\mathrm{FUV_{obs}}$,][]{Kang09} FUV magnitudes
    (proportional to the log of the modeled-to-observed flux ratio) versus
    deprojected area. The dashed line indicates where the magnitudes are equal.
    On average, the $\mathrm{FUV_{SFH}}$ values are $0.5\,\mathrm{mag}$ fainter
    than $\mathrm{FUV_{obs}}$ (dotted line) and the $1\sigma$ scatter is $\sim
    0.8\,\mathrm{mag}$. The systematic deficiency of the modeled magnitudes is
    likely due to overestimates in the FUV extinction. The scatter is greatest
    among the smallest regions and indicates that discrete sampling of the IMF
    is important on these scales. The combined region from Figure
    \ref{fig:cmd_grid}, indicated by the red square, shows much better
    agreement between $\mathrm{FUV_{SFH}}$ and $\mathrm{FUV_{obs}}$ than the
    individual regions it comprises.
}
\label{fig:fluxmod}
\end{figure}

% First mention: Figure 6
We also modeled the reddened FUV and NUV fluxes using the extinction model
described in \S \ref{sfhs.extinction}, the best-fit $A_{\mathrm{V}f}$ and
$dA_\mathrm{V}$ values in Table \ref{tab:SFHdata}, and the \citet[][see \S
\ref{discussion.fuv}]{Cardelli89} extinction curve. These fluxes were converted
into AB magnitudes and the uncertainties were evaluated in the same manner as
the intrinsic fluxes. We list the reddened FUV magnitudes,
$\mathrm{FUV_{SFH}}$, and the reddened UV colors, $\mathrm{(FUV-NUV)_{SFH}}$,
in Table \ref{tab:fluxmod}, and plot the difference between
$\mathrm{FUV_{SFH}}$ and $\mathrm{FUV_{obs}}$ versus deprojected region area in
Figure \ref{fig:fluxmod}. The comparison between the modeled and observed FUV
magnitudes is discussed in \S \ref{discussion.fuv}.

% Table 3
\input{table3}

% Figure 6

\section{SFR estimates}\label{sfrs}

The usual procedure for converting FUV flux into a SFR is to correct the
observed flux for extinction, calculate the luminosity, and then apply the
proper calibration. To test this method, we derived FUV extinction corrections,
$A_\mathrm{FUV}$, from the differences between $\mathrm{FUV_{SFH}}$ and
$\mathrm{FUV_{SFH,0}}$. The uncertainties in $A_\mathrm{FUV}$ were calculated
as described in \S \ref{sfhs.uncertainties}. The resulting values, listed in
Table \ref{tab:fluxmod}, were used to correct $\mathrm{FUV_{obs}}$.

% First mention: Table 4
The extinction-corrected observed FUV magnitudes were converted into SFRs,
$\mathrm{SFR_{FUV}}$, using the flux calibration from \citet{Kennicutt98} with
updated coefficients by \citet{Hao11} and \citet{Murphy11}
\citep[see][]{Kennicutt12}:
\begin{equation}
\left(\frac{\mathrm{SFR_{FUV}}}{M_\odot\,\mathrm{yr^{-1}}}\right) = 10^{-43.35}\,\left(\frac{L_{\mathrm{FUV}}}{\mathrm{erg\,s^{-1}}}\right)
\end{equation}
where $L_{\mathrm{FUV}}$ is the FUV luminosity in $\mathrm{erg\,s^{-1}}$. This
calibration was derived using the stellar population synthesis code Starburst99
\citep{Leitherer99} assuming that the SFR has been constant over the last
$100\,\mathrm{Myr}$. It also assumes a fully populated Kroupa IMF
\citep{Kroupa01} and solar metallicity.

The total uncertainties are the quadrature sum of the photometric uncertainties
propagated from $\mathrm{FUV_{obs}}$ and the random and systematic
uncertainties derived according to \S \ref{sfhs.uncertainties} (where
$\mathrm{SFR_{FUV}}$ was calculated for each value of $A_\mathrm{FUV}$ from the
HMC and $1\sigma$ SFHs). The $\mathrm{SFR_{FUV}}$ values are listed in Table
\ref{tab:fluxderived} and are shown against the SFHs in Figure
\ref{fig:sfh_grid}. Although more sophisticated tracers exist for calculating
SFRs \citep[e.g., hybrid tracers discussed in][]{Leroy12}, none of them can be
used with GALEX FUV and NUV data alone and such calculations are therefore
outside the scope of this study.

\begin{figure*}
\centering
\includegraphics[width=\textwidth]{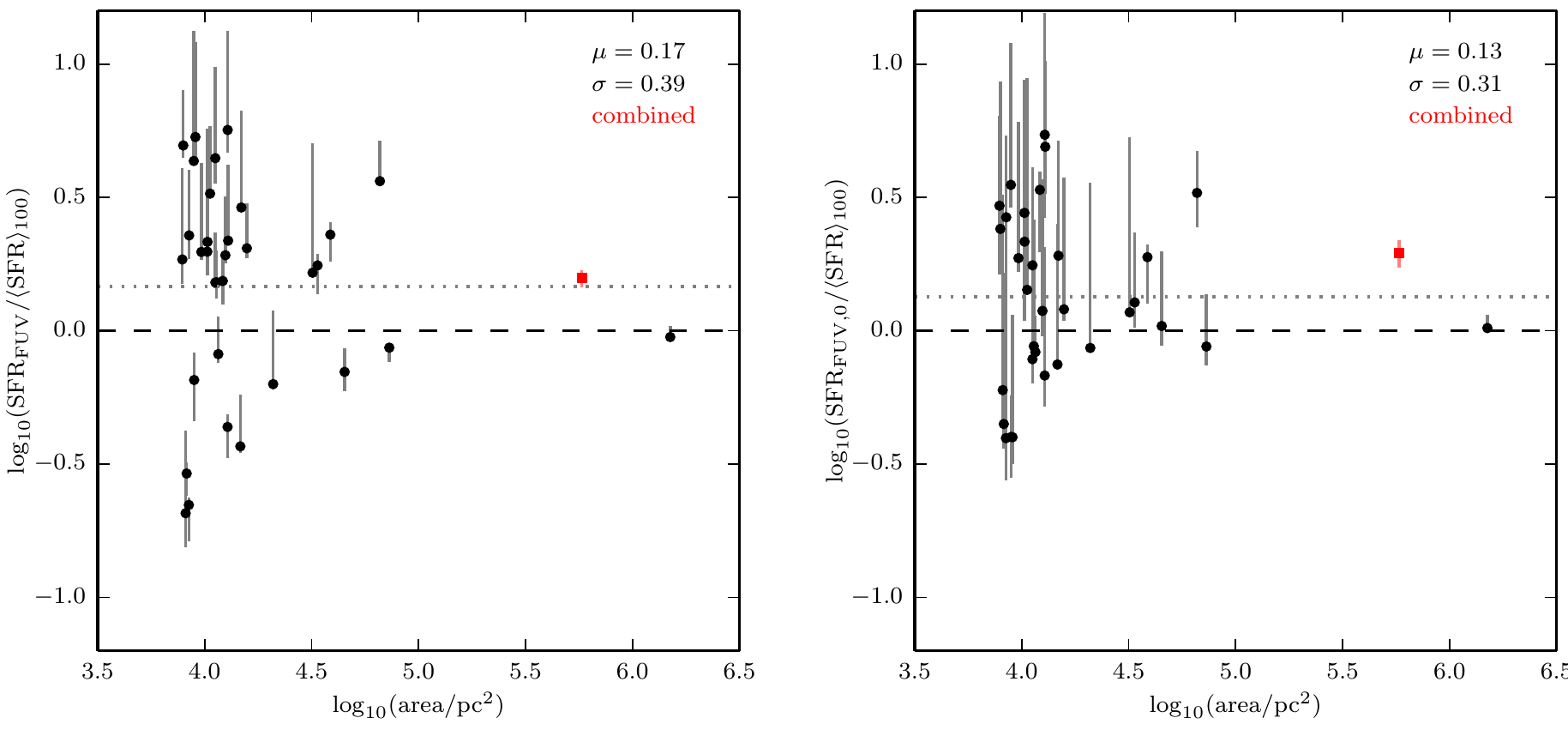}
\caption{The log ratio of the FUV flux-based SFRs to the $100\,\mathrm{Myr}$
    mean SFR of the SFHs ($\langle \mathrm{SFR}\rangle_{100}$) versus
    deprojected region area. In the left panel, the flux-based SFRs
    ($\mathrm{SFR_{FUV}}$) were derived from the extinction-corrected observed
    FUV magnitudes, and the flux-based SFRs in the right panel
    ($\mathrm{SFR_{FUV,0}}$) were derived from the intrinsic FUV magnitudes
    modeled from the SFHs. The dashed lines indicate where the SFRs are equal,
    and the mean and standard deviation of each log ratio is shown in its
    respective panel. The systematic overestimation of $\mathrm{SFR_{FUV}}$ in
    the left panel is likely due to overestimates in the FUV extinction, and
    the primary sources of the scatter are discrete sampling of the IMF and
    variability in the SFHs. The mean offset and the scatter are reduced for
    $\mathrm{SFR_{FUV,0}}$ in the right panel, where the remaining scatter is
    primarily due to SFH variability. In both panels, the combined region from
    Figure \ref{fig:cmd_grid}, indicated by the red square, shows much better
    agreement between the flux-based and mean SFRs than the individual regions
    it comprises.
}
\label{fig:sfrs}
\end{figure*}

% First mention: Figure 7
To compare with the flux-based SFRs, we calculated the mean SFR over the last
$100\,\mathrm{Myr}$ of the SFH, $\langle\mathrm{SFR}\rangle_{100} = M_{100}
\times 10^{-8}\,\mathrm{yr}^{-1}$, where $M_{100}$ is the total mass formed
over the same time period. The $\langle\mathrm{SFR}\rangle_{100}$ values are
shown in Figure \ref{fig:sfh_grid}, and both $M_{100}$ and
$\langle\mathrm{SFR}\rangle_{100}$ are listed in Table \ref{tab:SFHdata}. The
uncertainties were derived as described in \S \ref{sfhs.uncertainties}. We also
calculated the mean SFR over the last $500\,\mathrm{Myr}$, $\langle
\mathrm{SFR}\rangle_{500}$ (Figure \ref{fig:sfh_grid}). The $500\,\mathrm{Myr}$
timescale (the practical age limit of the SFHs, \S \ref{sfhs.results}) is
useful for understanding the overall behavior of the regions and illustrates
the significance of the SF activity in the last $100\,\mathrm{Myr}$ with
respect to the broader history. Figure \ref{fig:sfrs} shows the log ratio of
$\mathrm{SFR_{FUV}}$ to $\langle\mathrm{SFR}\rangle_{100}$ versus deprojected
region area, and is discussed in \S \ref{discussion.sfrs}.

% Table 4
\input{table4}

% Figure 7

\section{Discussion}\label{discussion}

\subsection{FUV magnitudes}\label{discussion.fuv}

The differences between the reddened FUV magnitudes modeled from the SFHs and
the observed FUV magnitudes of the regions, $\mathrm{FUV_{SFH} - FUV_{obs}}$,
shown in Figure \ref{fig:fluxmod} are normally distributed with a mean and
standard deviation of $\mu=0.09\,\mathrm{mag}$ and $\sigma=0.8\,\mathrm{mag}$,
respectively. The $\mathrm{FUV_{SFH}}$ values are consistent with the
$\mathrm{FUV_{obs}}$ values on average, demonstrating that the FUV magnitudes
are largely free of several potential systematic effects, such as scattering of
FUV photons from or into the regions or misinterpretation of the CMDs by MATCH.

The consistency of $\mu$ with zero supports the hypothesis from \S
\ref{sfhs.extinction} that the $A_{\mathrm{V},f},dA_\mathrm{V}$ extinction
model adequately describes the dust affecting the MS stars in the regions.
Because $A_\mathrm{FUV}$ is derived by extrapolating $A_{\mathrm{V},f}$ and
$dA_\mathrm{V}$ along an extinction curve, the lack of a significant offset
between $\mathrm{FUV_{SFH}}$ and $\mathrm{FUV_{obs}}$ justifies our adoption of
the average Galactic extinction curve from \citet{Cardelli89} with the standard
value of $R_\mathrm{V} = 3.1$. This is consistent with results from
\citet{Barmby00} and \citet{Bianchi96}, who found that the overall extinction
curves of M31 and the Galaxy are similar for optical and UV wavelengths,
respectively.

It is somewhat surprising that assuming the \citet{Cardelli89} extinction curve
does not produce a larger systematic offset between $\mathrm{FUV_{SFH}}$ and
$\mathrm{FUV_{obs}}$. Previous studies have shown that local dust properties
and the shape of the extinction curve strongly depend on environment
\citep{Fitzpatrick07, Bianchi11, Efremova11}, which brings into question the
applicability of any galaxy-averaged extinction curve to specific locations
within a galaxy. Furthermore, results from \citet{Bianchi11} and
\citet{Efremova11} indicate that areas of intense SF, such as UV-bright
regions, tend to have extinction curves that are steeper in the UV regime.
Despite these details, we find that, given a mean visual extinction of
$A_{\mathrm{V},f} + dA_\mathrm{V} \approx 1.5$, the $\mu=0.09\,\mathrm{mag}$
offset is consistent with a value of $R_\mathrm{V}$ between 3.1 and 3.2. This
is well within the range of $R_\mathrm{V}$ values obtained by
\citet{Fitzpatrick07} for 328 lines of sight in the Galaxy, for which the mean
and standard deviation was 3.0 and 0.3, respectively.

Although the modeled and the observed FUV magnitudes agree on average, Figure
\ref{fig:fluxmod} shows that the scatter in $\mathrm{FUV_{SFH}-FUV_{obs}}$
about the mean is larger than the uncertainties. A possible source of this
scatter is the assumption of a homogeneous metallicity for the modeled FUV
magnitudes, $[\mathrm{M/H}] = -0.3\,\mathrm{dex}$, whereas the actual final
metallicity values for most of the regions varied between $[\mathrm{M/H}] =
-0.6$ and $0.0\,\mathrm{dex}$ (\S \ref{sfhs.results}). Figure 6 in
\citet{Johnson13} shows how the FUV luminosity of a constant SFR model changes
as a function of input metallicity. Near $[\mathrm{M/H}] = -0.3\,\mathrm{dex}$
\citep[$\log_{10}(Z/Z_\odot) \approx -0.3$, assuming the helium-to-metals
enrichment law from][]{Bressan12}, changing $[\mathrm{M/H}]$ by $\pm
0.1\,\mathrm{dex}$ causes the modeled FUV flux to change by $\mp
0.015\,\mathrm{dex}$, or $\pm 0.038\,\mathrm{mag}$ in terms of FUV magnitudes.
Given a metallicity dispersion of $0.3\,\mathrm{dex}$, variations from the
assumed metallicity therefore lead to an uncertainty of about
$0.1\,\mathrm{mag}$ in $\mathrm{FUV_{SFH}}$. The effect of assuming a
homogeneous metallicity contributes only a small amount to the total scatter.

Figure \ref{fig:fluxmod} shows that the scatter in
$\mathrm{FUV_{SFH}-FUV_{obs}}$ appears to increase with decreasing region area.
Because the regions are all defined to have the same minimum FUV surface
brightness, the masses of the regions roughly scale with area, implying that
the scatter is greatest for the regions with the lowest masses. A well-known
characteristic of low-mass systems is that the distribution of stellar masses
is noticeably discrete, particularly with respect to the high-mass end of the
IMF where the relative probability of star formation is low. As a result, the
sampling of stellar masses from the IMF is not as complete in such systems as
for higher-mass systems. This is illustrated in the CMDs in Figure
\ref{fig:cmd_grid}, which show the upper MS in many of the regions to be
sparsely populated compared with the much larger region 4308.

To model UV fluxes from the SFHs, FSPS assumes that the stellar mass formed in
each age bin represents a full sampling of the IMF, which is inconsistent with
the actual sampling of stellar masses in the regions. Therefore, the modeled
flux is underestimated in regions that have an apparent excess of massive MS
stars relative to the number expected from a fully populated IMF, and is
overestimated in regions with an apparent lack of massive MS stars. The size of
this discrepancy should be larger for regions with lower masses due to the
sampling of the IMF becoming more discrete.\footnote{This effect is often
associated with the term, ``stochasticity''.} Given that area is a proxy for
mass in our sample, the scatter in Figure \ref{fig:fluxmod} is indeed
consistent with this expectation. We therefore consider the scatter in the
magnitudes to be caused by the application of the full-IMF assumption where the
effect of discrete sampling is important.

To further test the impact of region size on the magnitude discrepancy, we
constructed a larger effective region by combining the photometry of all
regions, excluding region 4308 (the largest region). We then measured the SFH
and modeled the total FUV magnitude following the same procedure used for the
other regions. The total area and effective observed FUV magnitude (from the
combination of the observed magnitudes of the individual regions) are given in
Table \ref{tab:observations}, and the CMD is shown in Figure
\ref{fig:cmd_grid}. We show the SFH and the corresponding best-fit extinction
parameters in Figure \ref{fig:sfh_grid} and Table \ref{tab:SFHdata},
respectively. The modeled FUV magnitude is listed in Table \ref{tab:fluxmod}.

The combined region in Figure \ref{fig:fluxmod} has a magnitude difference
similar to region 4308 and is more consistent with the sample mean than the
majority of the individual regions it comprises. The combined region apparently
produces a much better representation of stellar masses in the IMF than when
the regions are considered individually, making the combined region more
consistent with the full-IMF assumption. This result supports our hypothesis
that the scatter in the magnitudes is largely explained as a sampling effect of
the IMF.

We estimate that discrete sampling becomes important for the UV-bright regions
below an area of $\sim 10^5\,\mathrm{pc^2}$, and amounts to an uncertainty of
$\sigma=0.8\,\mathrm{mag}$ in the modeled FUV magnitudes, or a factor of
$10^{|\sigma/-2.5|} = 2$ in flux. Determining a characteristic area threshold
from our sample is difficult, however, due to the lack of regions with areas
between $10^5$ and $10^6\,\mathrm{pc^2}$.

\subsection{SFR estimates from FUV flux}\label{discussion.sfrs}

Figure \ref{fig:sfrs} shows many of the same features as Figure
\ref{fig:fluxmod}, namely log-normally distributed ratios with a mean offset
and scatter that is largest among the smallest regions (see \S
\ref{discussion.fuv}). The log-normal distribution for the
$\mathrm{SFR_{FUV}/\langle SFR\rangle_{100}}$ values shown in Figure
\ref{fig:sfrs} has $\mu=0.2$ and $\sigma=0.4$.

The offset in Figure \ref{fig:sfrs} is less than the scatter. The FUV-based
SFRs are therefore consistent with the mean SFRs from the SFHs on average,
although the offset is somewhat larger relative to the scatter than in Figure
\ref{fig:fluxmod}. The consistency of the FUV magnitudes in Figure
\ref{fig:fluxmod} shows that the offset in the SFR ratios is not due to
scattering of FUV photons, misinterpretation of the CMDs by MATCH, a deficiency
in the extinction model, or an inaccurate extinction curve. Additionally, both
$\mathrm{SFR_{FUV}}$ and $\langle \mathrm{SFR}\rangle_{100}$ assume a timescale
of $100\,\mathrm{Myr}$, so the offset is also not due to inconsistent
timescales.

One difference between $\mathrm{SFR_{FUV}}$ and $\langle
\mathrm{SFR}\rangle_{100}$, however, is that the FUV flux calibration was
derived assuming solar metallicity. The FUV brightness of a stellar population
decreases with increasing metallicity (see \S \ref{discussion.fuv}), so the SFR
of a high-metallicity population would need to be greater than that of a
low-metallicity population with the same FUV flux and SFH. Specifically,
overestimating $[\mathrm{M/H}]$ by $0.1\,\mathrm{dex}$ causes the SFR to be
overestimated by $0.015\,\mathrm{dex}$.

Because nearly all of the final metallicities were subsolar, the majority of
the $\mathrm{SFR_{FUV}}$ values are overestimated to some degree. Solar
metallicity is higher than the mean final metallicity from MATCH by
$0.3\,\mathrm{dex}$, so the FUV-based SFRs are overestimated by about
$0.05\,\mathrm{dex}$ on average. Variations from the metallicity assumed by the
flux calibration therefore account for approximately one third of the $\mu=0.2$
offset in Figure \ref{fig:sfrs}. With no other obvious systematic effects at
work, we attribute the remaining offset to low-number statistics.

Like FSPS, the FUV flux calibration from \citet{Kennicutt98} assumes that the
IMF is fully populated, so discrete sampling of the IMF should produce a
similar amount of scatter in Figures \ref{fig:fluxmod} and \ref{fig:sfrs}.
Regions that are brighter for their mass than expected from the full-IMF
assumption will have their FUV-based SFRs overestimated, and regions that are
fainter than expected will have their SFRs underestimated. As in Figure
\ref{fig:fluxmod}, Figure \ref{fig:sfrs} shows that this discrepancy increases
with decreasing area. The scatter in the SFR ratios therefore appears
consistent with the application of the full-IMF assumption to low-mass regions.
By comparing the $\sigma$ parameters of the log-normal distributions in Figures
\ref{fig:fluxmod} and \ref{fig:sfrs} ($|0.8/-2.5|=0.3$ and 0.4, respectively),
however, we find that the SFR ratios are somewhat more scattered than the
magnitude differences. This suggests that there is an additional factor
contributing to the scatter.

In addition to the full-IMF assumption, the FUV flux calibration assumes a
constant SFR over at least the past $\sim 100\,\mathrm{Myr}$. It is clear from
the SFHs in Figure \ref{fig:cmd_grid} that none of the regions are consistent
with this assumption. To test how the inconsistency with the constant-SFR
assumption affects the SFR estimates, we used the FUV flux calibration to
derive another set of SFRs, $\mathrm{SFR_{FUV,0}}$, from the modeled intrinsic
magnitudes in Table \ref{tab:fluxmod}. Both FSPS and the flux calibration
assume a fully populated IMF, so $\mathrm{SFR_{FUV,0}}$ is determined
self-consistently. Also, despite the fact that the regions are largely
inconsistent with the full-IMF assumption, $\mathrm{\langle SFR\rangle_{100}}$
depends only on the total mass formed in the SFH, not on how the stellar masses
were sampled from the IMF. Therefore, any discrepancies between
$\mathrm{SFR_{FUV,0}}$ and $\mathrm{\langle SFR\rangle_{100}}$ beyond the
measured uncertainties are not due to discrete IMF sampling.

The log ratios of $\mathrm{SFR_{FUV,0}}$ to $\langle \mathrm{SFR}\rangle_{100}$
are shown versus deprojected region area in Figure \ref{fig:sfrs}. As for the
SFR ratios from the observed FUV magnitudes, we assumed a log-normal
distribution and calculated $\mu$ and $\sigma$ to be 0.1 and 0.3, respectively.
As expected, we find that the FUV-based SFRs are consistent with the mean SFRs
on average. The SFR ratios are widely scattered, indicating that the accuracy
of the FUV-based SFR estimates is strongly affected by variations in the SFHs.
In the extreme case of an SSP, \citet{Leroy12} found that FUV-based SFR
estimates are intrinsically scattered by a factor of $\sim 3$ to 4 ($\sigma
\approx 0.5$ to 0.6 in log space) due to uncertainty about the age of the SSP
within a $100\,\mathrm{Myr}$ timescale. The uncertainty we measure,
$\sigma=0.3$, is within the intrinsic limit, which we expect given that the
regions are more complex than SSPs (see \S \ref{discussion.ssp}).

The dependence of the flux-based SFRs on the SFHs is further illustrated by the
combined region in Figure \ref{fig:sfrs}. The
$\mathrm{SFR_{FUV,0}}/\mathrm{\langle SFR\rangle_{100}}$ ratio for the combined
region is similar to that of region 4308 (the largest region) and is more
consistent with unity than for most of the individual regions it comprises. By
combining the regions, many of the variations in the individual SFHs are
averaged out and the combined SFH is more constant by comparison.

Taken together, the uncertainties due to discrete sampling of the IMF and
variability in the SFHs account for the total amount of scatter in the
$\mathrm{SFR_{FUV,0}/\langle SFR\rangle_{100}}$ ratios, as shown from the
quadrature sum of the $\sigma$ values of the log-normals, $0.3^2 + 0.3^2
\approx 0.4^2$. The uncertainty components are also the same size,
demonstrating that satisfying the full-IMF assumption and satisfying the
constant-SFR assumption are equally important for obtaining precise SFR
estimates from the FUV flux calibration. Inconsistencies with the full-IMF and
constant-SFR assumptions appear to become important in UV-bright regions
smaller than $\sim 10^5\,\mathrm{pc^2}$. Assuming that one has a robust FUV
extinction correction), FUV-based SFRs estimated for regions smaller than this
limit are uncertain by a factor of about 2.5. We stress that this factor
represents the best case uncertainty, as the dust corrections for integrated UV
flux are often unclear and substantial. Our results are consistent with
warnings from \citet{Murphy11}, \citet{Kennicutt12}, and \citet{Leroy12} that
flux calibrations become problematic on sub-kpc scales.

Perhaps the most important assumptions behind the flux calibration method are
the assumed SFH and that SFR has a clear relationship with observed flux.
However, it is ambiguous whether the difference in flux between two populations
is due to a simple difference in SFR because the observed flux strongly depends
on the underlying SFH. This dependence is observed on scales both large and
small, e.g., as demonstrated for UV color in spiral galaxies by
\citet{Barnes11} and for FUV flux in our sample of UV-bright regions. We find,
for example, that although region 4299 and 4350 have total masses (and thus
mean SFRs) within 5\%, their SFHs are quite different and the FUV flux of
region 4350 is more than a factor of 2 brighter than that of 4299. We also find
that FUV flux is often degenerate with SFH (a wide range of SFHs can give rise
to the same FUV flux), such as the case for regions 4318 and 4330. These
complexities illustrate the inherent difficulty of using integrated flux alone
to characterize SF.

\subsection{SSP ages and masses}\label{discussion.ssp}

Because the integrated UV spectrum of an SSP evolves significantly over
relatively short timescales \citep[$\sim$ few Myr, as indicated in SSP models
from][]{Leitherer99}, the integrated $\mathrm{FUV-NUV}$ color can, in
principle, be used to estimate age through population synthesis modeling.
Clearly, this technique rests on the assumption that the population
approximates an SSP, i.e., that the population can be characterized by a single
age. The SSP assumption is typically acceptable for stellar clusters where
stars are generally formed at the same epoch, but it becomes untenable whenever
the SFH is more complex than a single SF event.

\citetalias{Kang09} estimated SSP ages, $\mathrm{Age_{SSP}}$, for the regions
by comparing the observed $\mathrm{FUV-NUV}$ color with Padova stellar
evolution models \citep{Girardi10} for a range of metallicities and dust types.
The models were reddened according to the average $E(\mathrm{B-V})$ measured in
the regions (see \S \ref{sfhs.extinction}). The ages derived for solar
metallicity and $R_\mathrm{V}=3.1$ are shown with the SFHs in Figure
\ref{fig:sfh_grid} and are listed in Table \ref{tab:fluxderived}. It is
immediately clear from the SFHs that the majority of the regions to not
approximate SSPs and that the very concept of assigning single ages to these
regions is invalid. Furthermore, the $\mathrm{Age_{SSP}}$ values often do not
correspond to the main episodes of SF.

\begin{figure}
\centering
\includegraphics[width=\columnwidth]{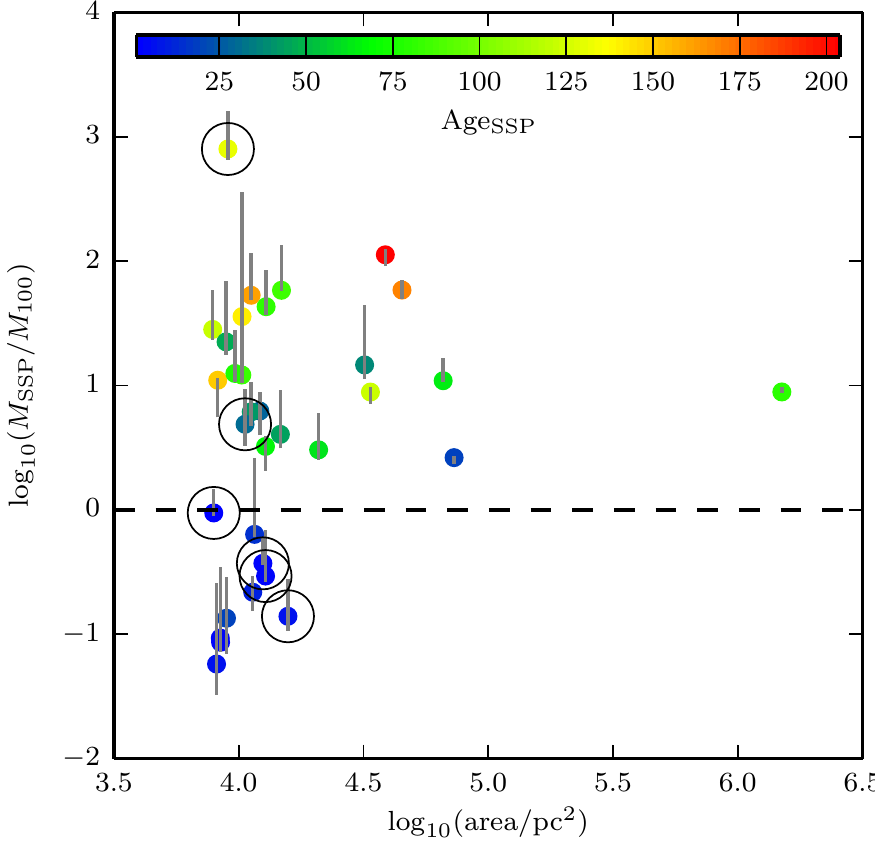}
\caption{The log ratio of the SSP masses based on UV color and luminosity
    ($M_\mathrm{SSP}$) to the total mass formed over the last
    $100\,\mathrm{Myr}$ of the SFHs ($M_{100}$) versus deprojected region area.
    The markers are colored according to the SSP age, $\mathrm{Age_{SSP}}$,
    corresponding to $M_\mathrm{SSP}$. The dashed line indicates where the
    masses are equal. SSP masses are underestimated and overestimated for
    regions that are estimated to be young and old, respectively. The circled
    points indicate the most SSP-like regions identified in the sample \S
    \ref{discussion.ssp}. At $\log_{10}(M_\mathrm{SSP}/M_{100}) \approx 3$,
    region 4331 is extremely discrepant and we do not include it in our SSP
    analysis. The $M_\mathrm{SSP}$ values for the other SSP-like regions
    indicate a factor of 3 to 4 uncertainty with respect to the $M_{100}$
    values.
}
\label{fig:masses}
\end{figure}

% First mention: Figure 8
The corresponding SSP masses, $M_\mathrm{SSP}$, were estimated by
\citetalias{Kang09} from $\mathrm{Age_{SSP}}$ and the FUV luminosity. We list
the $M_\mathrm{SSP}$ values in Table \ref{tab:fluxderived} and show the log of
$M_\mathrm{SSP}/M_{100}$ versus deprojected region area in Figure
\ref{fig:masses}. Most of the $M_\mathrm{SSP}$ values are within one to two
orders of magnitude of $M_{100}$. This large uncertainty range is a consequence
of applying the SSP assumption to regions that are generally not SSPs. By
coloring each point in Figure \ref{fig:masses} by $\mathrm{Age_{SSP}}$, we find
that the masses for regions determined to be young are underestimated, and the
masses for regions determined to be older are overestimated (an old SSP must be
more massive than a young SSP to have the same UV luminosity). This trend is
observed independent of region size.

% Figure 8

Although most of the regions are not SSPs, we do find that the SSP assumption
is justified in some cases. To identify SSP-like regions, we calculated the
ratio of the mass formed in the age bin with the highest SFR over the last
$100\,\mathrm{Myr}$, $M_\mathrm{peak}$, to the total mass formed over the same
time period, $M_{100}$. The values of $M_\mathrm{peak}$,
$M_\mathrm{peak}/M_{100}$, and $\mathrm{Age_{peak}}$ (the mean age of the bin
containing $M_\mathrm{peak}$) are listed in Table \ref{tab:SFHdata}. We
considered any region with $M_\mathrm{peak}/M_{100} \ge 0.9$ to be consistent
with an SSP, i.e., any region that has formed more than 90\% of its total mass
in a single age bin over the last $100\,\mathrm{Myr}$. We found that 18\% of
the regions (6 of 33; 4331, 4333, 4345, 4355, 4360, and 4362) meet this
criterion and we indicate them in Tables \ref{tab:observations},
\ref{tab:SFHdata}, \ref{tab:fluxmod} and \ref{tab:fluxderived}.

Except for region 4331, the mean discrepancy between $\mathrm{Age_{SSP}}$ and
$\mathrm{Age_{peak}}$ for the SSP-like regions is $10\,\mathrm{Myr}$. On
average, $M_\mathrm{SSP}$ is consistent with $M_{100}$ to within a factor of 3
or 4 (excluding region 4331). These age and mass discrepancies are often larger
than the error bars shown in Figures \ref{fig:sfh_grid} and \ref{fig:masses}.
Region 4331 is extremely discrepant in both age ($\mathrm{Age_{SSP} -
Age_{peak}} = 80\,\mathrm{Myr}$) and mass ($M_\mathrm{SSP}/M_{100}=800$) for
reasons that are unclear, and we do not include it in the remaining analyses.

The uncertainties in $\mathrm{Age_{SSP}}$ are about $3\,\mathrm{Myr}$, on
average, and are propagated solely from the photometric uncertainties. We
suggest that these uncertainties are underreported, given that metallicity and
extinction can potentially introduce systematics that are much larger than the
uncertainties in FUV luminosity and $\mathrm{FUV-NUV}$ color. We consider below
how metallicity and extinction affect the age and mass estimates.

Metallicity increases the rate at which an SSP becomes redder with age
\citep{Bianchi11, Kang09}. Therefore, if metallicity is overestimated, then
$\mathrm{Age_{SSP}}$ will be underestimated. To test how significantly
metallicity affects the ages of the SSP-like regions, we compared the ages for
solar metal abundance with the ages derived by \citetalias{Kang09} for
$Z=0.05$. The change in metallicity decreased the SSP age estimates by an
average of $6\,\mathrm{Myr}$. We also compared the corresponding masses and
found that the masses for solar metal abundance were, on average, a factor of 2
larger than the those for $Z=0.05$.

Since UV flux is highly susceptible to dust extinction, any errors in the
applied reddening values can significantly affect the derived SSP ages and
masses. Namely, if $E(\mathrm{B-V})$ is underestimated, then the corrected
$\mathrm{FUV-NUV}$ color will be redder than it should be \citep[assuming the
extinction curve of][although this is another potential source of
uncertainty]{Cardelli89}, causing an overestimate in $\mathrm{Age_{SSP}}$. The
change in $M_\mathrm{SSP}$ is less clear because underestimating the FUV
luminosity and overestimating the age have opposite effects.

To quantify the impact of changes in $\mathrm{FUV-NUV}$ color on
$\mathrm{Age_{SSP}}$ and $M_\mathrm{SSP}$, we used FSPS (see \S \ref{fluxmod})
to model the time evolution of the UV magnitudes of an SSP with solar
metallicity. For a $10\,\mathrm{Myr}$ old SSP ($\mathrm{FUV-NUV} \approx
-0.09$), we found that a $0.1\,\mathrm{mag}$ reduction in $E(\mathrm{B-V})$
causes the age to be overestimated by $25\,\mathrm{Myr}$ and the mass to be
overestimated by a factor of 2.6. $\mathrm{Age_{SSP}}$ and $M_\mathrm{SSP}$ are
therefore highly sensitive to changes in $\mathrm{FUV-NUV}$.

Systematics from the assumed metallicity and the extinction correction can
plausibly account for the observed age and mass discrepancies among the
SSP-like regions. From these discrepancies, we propose that more realistic
uncertainties for the ages and the masses derived from UV flux are
$10\,\mathrm{Myr}$ and a factor of 3 to 4, respectively, though the limited
number of regions resembling SSPs makes these uncertainties difficult to
determine more precisely. The most striking aspect of this analysis is that
over 80\% of the regions in the sample are entirely inconsistent with the SSP
assumption in the first place, a fact that could not be known without measuring
the SFHs, calling into question the practice of deriving ages and masses for
populations that are not confirmed SSPs.

\section{Conclusion}\label{conclusion}

In this study, we have derived the recent ($< 500\,\mathrm{Myr}$) SFHs of 33
UV-bright regions in M31 using optical HST observations from PHAT. The regions
were defined by \citetalias{Kang09} based on GALEX FUV surface brightness and
have areas ranging from $8 \times 10^3$ to $1.5 \times 10^6\,\mathrm{pc}^2$. We
used the SFH code MATCH to fit the CMDs of the regions and measure their the
SFHs based on the resolved stars from the PHAT photometry. We modeled the
extinction in the regions using a foreground parameter and a differential
parameter, which were optimized for each region to find the best-fit SFH.

We used FSPS to model both the intrinsic and reddened FUV and NUV magnitudes of
the regions based on their SFHs. The differences between the modeled reddened
and the observed FUV magnitudes, $\mathrm{FUV_{SFH}-FUV_{obs}}$, followed a
normal distribution with $\mu=0.09$ and $\sigma=0.3$. On average, the
$\mathrm{FUV_{SFH}}$ values were consistent with the $\mathrm{FUV_{obs}}$
values, confirming the reliability of the SFHs, our extinction model, and the
\citet{Cardelli89} extinction curve. We attribute the scatter in the flux
ratios to the assumption made by FSPS that the IMF is fully populated while the
actual distribution of stellar masses becomes more discrete as smaller regions
are considered.

The observed, extinction-corrected FUV magnitudes were converted into SFRs,
$\mathrm{SFR_{FUV}}$, using the FUV flux calibration from \citet{Kennicutt98}
with updated coefficients by \citet{Hao11} and \citet{Murphy11}. We also
derived the mean SFRs for the last $100\,\mathrm{Myr}$ of the SFHs, $\langle
\mathrm{SFR}\rangle_{100}$. The $\mathrm{SFR_{FUV}}/\langle
\mathrm{SFR}\rangle_{100}$ ratios were log-normally distributed with $\mu=0.2$
and $\sigma=0.4$. Overall, the $\mathrm{SFR_{FUV}}$ values were consistent with
the $\langle \mathrm{SFR}\rangle_{100}$ values, though a small amount of the
offset was attributable to inconsistencies with the metallicity assumed by the
flux calibration.

The intrinsic modeled FUV magnitudes were also converted into SFRs,
$\mathrm{SFR_{FUV,0}}$, which were free from biases due to extinction
corrections and IMF sampling. The log-normal for the
$\mathrm{SFR_{FUV,0}}/\langle \mathrm{SFR}\rangle_{100}$ ratios had $\mu=0.1$
and $\sigma=0.3$, indicating that assuming a constant SFR (implicit in the flux
calibration) for regions with highly variable SFHs is an important source of
scatter. We conclude that the total scatter in the $\mathrm{SFR_{FUV}}/\langle
\mathrm{SFR}\rangle_{100}$ ratio is due to the assumptions of a full IMF and a
constant SFR in regions where discrete sampling of the IMF and high variability
in the SFHs are important. Combined, these effects result in a factor of 2.5
uncertainty in the FUV-based SFRs. Although there is a significant lack of
regions in our sample with areas between $10^5$ and $10^6\,\mathrm{pc}^2$, we
estimate that discrete IMF sampling and SFH variability become important below
$10^5\,\mathrm{pc}^2$, or scales of a few hundred pc.

Ages and masses were derived for the regions by \citetalias{Kang09} from
observed $\mathrm{FUV-NUV}$ color and FUV luminosity, using the assumption that
the regions are SSPs. By comparing the ages to the SFHs, we found that most of
the regions are entirely inconsistent with the SSP assumption. Furthermore, the
ages often did not correspond to the main episodes of SF, and the masses were
discrepant with the masses integrated from the SFHs by up to 2 orders of
magnitude. These results call into question the practice of deriving ages and
masses for populations that are not confirmed SSPs.

We identified SSP-like regions as regions which formed 90\% or more of their
mass over the past $100\,\mathrm{Myr}$ in a single age bin of their SFH. These
regions accounted for 18\% of our sample (6 of 33). Among this subset, we found
discrepancies of $10\,\mathrm{Myr}$ in the ages and a factor of $3-4$ in the
masses derived from UV flux, most likely due to systematics in metallicity and
extinction. We propose that these discrepancies represent realistic
uncertainties in the SSP ages and masses, though the limited number of SSP-like
regions in our sample makes the uncertainties difficult to determine. Finally,
identification of the SSP-like regions was not possible from integrated FUV
flux.

We thank Y.~Kang for his assistance in understanding the results presented in
\citet{Kang09}. This research has made use of NASA's Astrophysics Data System
Bibliographic Services and the NASA/IPAC Extragalactic Database (NED), which is
operated by the Jet Propulsion Laboratory, California Institute of Technology,
under contract with the National Aeronautics and Space Administration. This
work was supported by the Space Telescope Science Institute through GO-12055.
Support for D.~R.~W is provided by NASA through Hubble Fellowship grant
HST-HF-51331.01 awarded by the Space Telescope Science Institute, which is
operated by the Association of Universities for Research in Astronomy, Inc.,
under NASA contract NAS 5-26555. This research made use of Astropy, a
community-developed core Python package for Astronomy \citep{astropy}, as well
as NumPy and SciPy \citet{scipy}, IPython \citep{ipython}, and Matplotlib
\citep{matplotlib}.

\end{document}

%% file: table1.tex
\begin{deluxetable*}{ccccccc}
\tablecaption{Observational properties of UV-bright regions in PHAT Brick 15.
    \label{tab:observations}
}
\tablewidth{0pt}
\tablehead{
    \colhead{ID} &
    \colhead{RA\tablenotemark{a}} &
    \colhead{dec\tablenotemark{a}} &
    \colhead{area} &
    \colhead{area\tablenotemark{b}} &
    \colhead{$\mathrm{FUV_{obs}}$\tablenotemark{a}} &
    \colhead{$\mathrm{(FUV-NUV)_{obs}}$\tablenotemark{a}} \\
    \colhead{} &
    \colhead{(deg)} &
    \colhead{(deg)} &
    \colhead{($10^2\,\mathrm{arcsec}^2$)} &
    \colhead{($10^3\,\mathrm{pc}^2$)} &
    \colhead{(AB mag)} &
    \colhead{(AB mag)}
}
\startdata
4285\phantom{*} &  11.352559 &  41.824921 &  \phantom{0}\phantom{0}1.6 &  \phantom{0}\phantom{0}11.2 &  $18.570\pm 0.017$ &  $\phantom{+}0.325\pm 0.019$ \\
4288\phantom{*} &  11.352191 &  41.830040 &  \phantom{0}\phantom{0}1.5 &  \phantom{0}\phantom{0}10.3 &  $19.281\pm 0.025$ &  $\phantom{+}0.147\pm 0.029$ \\
4290\phantom{*} &  11.364936 &  41.833477 &  \phantom{0}\phantom{0}5.6 &  \phantom{0}\phantom{0}38.7 &  $17.377\pm 0.010$ &  $\phantom{+}0.468\pm 0.011$ \\
4292\phantom{*} &  11.120670 &  41.833038 &  \phantom{0}\phantom{0}1.2 &  \phantom{0}\phantom{0}\phantom{0}8.4 &  $20.457\pm 0.052$ &  $-0.120\pm 0.064$ \\
4293\phantom{*} &  11.108700 &  41.837337 &  \phantom{0}\phantom{0}6.5 &  \phantom{0}\phantom{0}45.1 &  $18.160\pm 0.016$ &  $\phantom{+}0.388\pm 0.018$ \\
4298\phantom{*} &  11.345233 &  41.845989 &  \phantom{0}\phantom{0}4.9 &  \phantom{0}\phantom{0}33.7 &  $17.859\pm 0.013$ &  $\phantom{+}0.253\pm 0.015$ \\
4299\phantom{*} &  11.123035 &  41.843586 &  \phantom{0}\phantom{0}3.0 &  \phantom{0}\phantom{0}20.9 &  $19.319\pm 0.030$ &  $\phantom{+}0.087\pm 0.035$ \\
4308\phantom{*} &  11.152348 &  41.874954 &  216.5 &  1502.3 &  $13.898\pm 0.002$ &  $\phantom{+}0.181\pm 0.002$ \\
4310\phantom{*} &  11.197878 &  41.852535 &  \phantom{0}\phantom{0}1.2 &  \phantom{0}\phantom{0}\phantom{0}8.2 &  $19.411\pm 0.027$ &  $\phantom{+}0.303\pm 0.030$ \\
4313\phantom{*} &  11.234838 &  41.853275 &  \phantom{0}\phantom{0}1.5 &  \phantom{0}\phantom{0}10.3 &  $19.026\pm 0.022$ &  $\phantom{+}0.290\pm 0.025$ \\
4314\phantom{*} &  11.082835 &  41.854801 &  \phantom{0}\phantom{0}1.8 &  \phantom{0}\phantom{0}12.2 &  $19.714\pm 0.032$ &  $\phantom{+}0.115\pm 0.037$ \\
4317\phantom{*} &  11.216114 &  41.862221 &  \phantom{0}\phantom{0}1.1 &  \phantom{0}\phantom{0}\phantom{0}7.9 &  $19.556\pm 0.029$ &  $\phantom{+}0.259\pm 0.033$ \\
4318\phantom{*} &  11.218816 &  41.869392 &  \phantom{0}\phantom{0}1.3 &  \phantom{0}\phantom{0}\phantom{0}8.9 &  $19.978\pm 0.038$ &  $-0.058\pm 0.045$ \\
4320\phantom{*} &  11.325653 &  41.868969 &  \phantom{0}\phantom{0}2.1 &  \phantom{0}\phantom{0}14.8 &  $18.200\pm 0.014$ &  $\phantom{+}0.195\pm 0.016$ \\
4321\phantom{*} &  11.206724 &  41.872875 &  \phantom{0}\phantom{0}1.8 &  \phantom{0}\phantom{0}12.8 &  $19.823\pm 0.037$ &  $\phantom{+}0.118\pm 0.043$ \\
4322\phantom{*} &  11.193023 &  41.873569 &  \phantom{0}\phantom{0}1.2 &  \phantom{0}\phantom{0}\phantom{0}8.2 &  $20.810\pm 0.067$ &  $-0.164\pm 0.085$ \\
4330\phantom{*} &  11.244569 &  41.897583 &  \phantom{0}\phantom{0}2.1 &  \phantom{0}\phantom{0}14.7 &  $19.848\pm 0.039$ &  $\phantom{+}0.087\pm 0.046$ \\
4331\tablenotemark{*} &  11.343492 &  41.897060 &  \phantom{0}\phantom{0}1.3 &  \phantom{0}\phantom{0}\phantom{0}9.1 &  $18.411\pm 0.016$ &  $\phantom{+}0.462\pm 0.017$ \\
4333\tablenotemark{*} &  11.086989 &  41.904243 &  \phantom{0}\phantom{0}1.8 &  \phantom{0}\phantom{0}12.5 &  $18.811\pm 0.019$ &  $-0.271\pm 0.023$ \\
4335\phantom{*} &  11.165060 &  41.908730 &  \phantom{0}\phantom{0}4.6 &  \phantom{0}\phantom{0}32.0 &  $18.172\pm 0.016$ &  $\phantom{+}0.026\pm 0.018$ \\
4337\phantom{*} &  11.245595 &  41.910343 &  \phantom{0}\phantom{0}1.7 &  \phantom{0}\phantom{0}11.6 &  $19.294\pm 0.026$ &  $\phantom{+}0.014\pm 0.031$ \\
4339\phantom{*} &  11.125310 &  41.918499 &  \phantom{0}\phantom{0}9.5 &  \phantom{0}\phantom{0}66.0 &  $16.591\pm 0.007$ &  $\phantom{+}0.087\pm 0.008$ \\
4345\tablenotemark{*} &  11.103488 &  41.922085 &  \phantom{0}\phantom{0}2.3 &  \phantom{0}\phantom{0}15.8 &  $18.625\pm 0.018$ &  $-0.182\pm 0.022$ \\
4346\phantom{*} &  11.244633 &  41.928699 &  \phantom{0}10.5 &  \phantom{0}\phantom{0}73.0 &  $17.194\pm 0.010$ &  $-0.007\pm 0.012$ \\
4348\phantom{*} &  11.261269 &  41.925659 &  \phantom{0}\phantom{0}1.6 &  \phantom{0}\phantom{0}11.4 &  $18.717\pm 0.019$ &  $-0.129\pm 0.022$ \\
4349\phantom{*} &  11.222815 &  41.925503 &  \phantom{0}\phantom{0}1.9 &  \phantom{0}\phantom{0}12.9 &  $19.313\pm 0.026$ &  $\phantom{+}0.248\pm 0.030$ \\
4350\phantom{*} &  11.230723 &  41.930325 &  \phantom{0}\phantom{0}1.6 &  \phantom{0}\phantom{0}11.2 &  $19.514\pm 0.030$ &  $\phantom{+}0.062\pm 0.034$ \\
4353\phantom{*} &  11.163334 &  41.939396 &  \phantom{0}\phantom{0}1.2 &  \phantom{0}\phantom{0}\phantom{0}8.5 &  $20.168\pm 0.040$ &  $-0.241\pm 0.049$ \\
4354\phantom{*} &  11.090720 &  41.946636 &  \phantom{0}\phantom{0}1.3 &  \phantom{0}\phantom{0}\phantom{0}8.9 &  $18.454\pm 0.016$ &  $\phantom{+}0.037\pm 0.018$ \\
4355\tablenotemark{*} &  11.197759 &  41.949593 &  \phantom{0}\phantom{0}1.1 &  \phantom{0}\phantom{0}\phantom{0}7.9 &  $19.008\pm 0.021$ &  $-0.289\pm 0.025$ \\
4360\tablenotemark{*} &  11.140469 &  41.954372 &  \phantom{0}\phantom{0}1.8 &  \phantom{0}\phantom{0}12.8 &  $18.533\pm 0.017$ &  $-0.256\pm 0.020$ \\
4362\tablenotemark{*} &  11.125412 &  41.955841 &  \phantom{0}\phantom{0}1.5 &  \phantom{0}\phantom{0}10.6 &  $19.363\pm 0.026$ &  $-0.010\pm 0.030$ \\
4364\phantom{*} &  11.133373 &  41.959290 &  \phantom{0}\phantom{0}1.4 &  \phantom{0}\phantom{0}\phantom{0}9.7 &  $19.342\pm 0.025$ &  $\phantom{+}0.122\pm 0.029$ \\
comb.\tablenotemark{c} &   &   &  \phantom{0}83.6 &  \phantom{0}580.3 &  $14.764\pm 0.003$ &  $\phantom{+}0.139\pm 0.004$
\enddata
\tablenotetext{a}{\citet{Kang09}. The magnitudes have not been corrected for
    extinction.
}
\tablenotetext{b}{Calculated from the solid angles (areas in
    $\mathrm{arcsec}^2$) assuming a distance of $785\,\mathrm{kpc}$
    \citep{McConnachie05} and deprojected assuming an inclination of
    $78\,\mathrm{deg}$ \citep{Tully94}.
}
\tablenotetext{c}{The combination of all regions except for region 4308.}
\tablenotetext{*}{SSP-like region (\S \ref{discussion.ssp}).}
\end{deluxetable*}

%% file: table2.tex
%%% For asymmetric uncertainties, `\substack{+x \\ -y}` from amsmath is
%%% slightly more compact than `^{+x}_{-y}`.

\begin{deluxetable*}{ccccccc}
\tablecaption{Region properties derived from the SFHs.\label{tab:SFHdata}}
\tablewidth{0pt}
\tablehead{
    \colhead{ID} &
    \colhead{$A_{\mathrm{V}f}$\tablenotemark{a}} &
    \colhead{$dA_\mathrm{V}$\tablenotemark{a}} &
    \colhead{$M_{100}$, $\langle \mathrm{SFR}\rangle_{100}$\tablenotemark{b}} &
    \colhead{$M_\mathrm{peak}$\tablenotemark{c}} &
    \colhead{$M_\mathrm{peak}/M_{100}$} &
    \colhead{$\mathrm{Age_{peak}}$\tablenotemark{d}} \\
    \colhead{} &
    \colhead{} &
    \colhead{} &
    \colhead{$(10^3\,M_\odot$, $10^{-5}\,M_\odot\,\mathrm{yr}^{-1})$} &
    \colhead{$(10^3\,M_\odot)$} &
    \colhead{} &
    \colhead{(Myr)}
}
\startdata
4285\phantom{*} &  $0.30\substack{+0.00 \\ -0.05}$ &  $2.50\substack{+0.05 \\ -0.25}$ &  $\phantom{0}\phantom{0}\phantom{0}3.0\substack{+2.4\phantom{0}\phantom{0} \\ -0.3\phantom{0}\phantom{0}}$ &  $\phantom{0}\phantom{0}\phantom{0}2.4\substack{+0.0\phantom{0}\phantom{0} \\ -1.8\phantom{0}\phantom{0}}$ &  0.78 &  $23.8\substack{+52.3 \\ -8.8\phantom{0}}$ \\[\dy]
4288\phantom{*} &  $0.60\substack{+0.00 \\ -0.00}$ &  $0.00\substack{+0.00 \\ -0.00}$ &  $\phantom{0}\phantom{0}\phantom{0}1.2\substack{+1.2\phantom{0}\phantom{0} \\ -0.0\phantom{0}\phantom{0}}$ &  $\phantom{0}\phantom{0}\phantom{0}0.6\substack{+0.0\phantom{0}\phantom{0} \\ -0.3\phantom{0}\phantom{0}}$ &  0.54 &  $\phantom{0}6.7\substack{+54.0 \\ -4.3\phantom{0}}$ \\[\dy]
4290\phantom{*} &  $0.40\substack{+0.00 \\ -0.05}$ &  $1.95\substack{+0.30 \\ -0.05}$ &  $\phantom{0}\phantom{0}17.8\substack{+1.8\phantom{0}\phantom{0} \\ -3.7\phantom{0}\phantom{0}}$ &  $\phantom{0}\phantom{0}\phantom{0}6.3\substack{+1.6\phantom{0}\phantom{0} \\ -3.7\phantom{0}\phantom{0}}$ &  0.35 &  $\phantom{0}6.0\substack{+112.2 \\ -0.7\phantom{0}}$ \\[\dy]
4292\phantom{*} &  $0.70\substack{+0.00 \\ -0.05}$ &  $1.10\substack{+0.05 \\ -0.05}$ &  $\phantom{0}\phantom{0}14.0\substack{+0.4\phantom{0}\phantom{0} \\ -3.7\phantom{0}\phantom{0}}$ &  $\phantom{0}\phantom{0}10.4\substack{+0.3\phantom{0}\phantom{0} \\ -5.7\phantom{0}\phantom{0}}$ &  0.74 &  $75.1\substack{+21.5 \\ -7.7\phantom{0}}$ \\[\dy]
4293\phantom{*} &  $0.40\substack{+0.00 \\ -0.00}$ &  $1.00\substack{+0.10 \\ -0.05}$ &  $\phantom{0}\phantom{0}15.4\substack{+2.9\phantom{0}\phantom{0} \\ -2.5\phantom{0}\phantom{0}}$ &  $\phantom{0}\phantom{0}\phantom{0}8.4\substack{+0.6\phantom{0}\phantom{0} \\ -5.7\phantom{0}\phantom{0}}$ &  0.55 &  $66.9\substack{+27.6 \\ -11.5}$ \\[\dy]
4298\phantom{*} &  $0.45\substack{+0.00 \\ -0.00}$ &  $1.95\substack{+0.05 \\ -0.25}$ &  $\phantom{0}\phantom{0}16.9\substack{+1.6\phantom{0}\phantom{0} \\ -3.7\phantom{0}\phantom{0}}$ &  $\phantom{0}\phantom{0}\phantom{0}9.4\substack{+0.4\phantom{0}\phantom{0} \\ -6.9\phantom{0}\phantom{0}}$ &  0.56 &  $47.4\substack{+5.7\phantom{0} \\ -45.2}$ \\[\dy]
4299\phantom{*} &  $0.35\substack{+0.00 \\ -0.00}$ &  $0.70\substack{+0.05 \\ -0.00}$ &  $\phantom{0}\phantom{0}\phantom{0}3.9\substack{+2.5\phantom{0}\phantom{0} \\ -0.1\phantom{0}\phantom{0}}$ &  $\phantom{0}\phantom{0}\phantom{0}2.2\substack{+1.1\phantom{0}\phantom{0} \\ -1.4\phantom{0}\phantom{0}}$ &  0.55 &  $94.6\substack{+0.0\phantom{0} \\ -42.1}$ \\[\dy]
4308\phantom{*} &  $0.50\substack{+0.00 \\ -0.00}$ &  $1.55\substack{+0.00 \\ -0.00}$ &  $1103.5\substack{+102.7 \\ -16.6\phantom{0}}$ &  $\phantom{0}208.7\substack{+49.9\phantom{0} \\ -37.3\phantom{0}}$ &  0.19 &  $29.9\substack{+43.5 \\ -1.5\phantom{0}}$ \\[\dy]
4310\phantom{*} &  $0.40\substack{+0.05 \\ -0.00}$ &  $0.45\substack{+0.00 \\ -0.10}$ &  $\phantom{0}\phantom{0}\phantom{0}7.0\substack{+0.2\phantom{0}\phantom{0} \\ -1.3\phantom{0}\phantom{0}}$ &  $\phantom{0}\phantom{0}\phantom{0}4.3\substack{+0.0\phantom{0}\phantom{0} \\ -3.1\phantom{0}\phantom{0}}$ &  0.61 &  $59.7\substack{+7.4\phantom{0} \\ -38.9}$ \\[\dy]
4313\phantom{*} &  $0.60\substack{+0.05 \\ -0.10}$ &  $0.00\substack{+0.15 \\ -0.00}$ &  $\phantom{0}\phantom{0}\phantom{0}1.3\substack{+1.2\phantom{0}\phantom{0} \\ -0.2\phantom{0}\phantom{0}}$ &  $\phantom{0}\phantom{0}\phantom{0}0.5\substack{+0.3\phantom{0}\phantom{0} \\ -0.3\phantom{0}\phantom{0}}$ &  0.38 &  $\phantom{0}6.7\substack{+56.9 \\ -2.0\phantom{0}}$ \\[\dy]
4314\phantom{*} &  $0.95\substack{+0.05 \\ -0.05}$ &  $1.50\substack{+0.10 \\ -0.05}$ &  $\phantom{0}\phantom{0}10.3\substack{+1.5\phantom{0}\phantom{0} \\ -1.8\phantom{0}\phantom{0}}$ &  $\phantom{0}\phantom{0}\phantom{0}4.0\substack{+0.4\phantom{0}\phantom{0} \\ -2.4\phantom{0}\phantom{0}}$ &  0.39 &  $\phantom{0}2.2\substack{+80.5 \\ -0.0\phantom{0}}$ \\[\dy]
4317\phantom{*} &  $0.45\substack{+0.15 \\ -0.10}$ &  $0.70\substack{+0.25 \\ -0.15}$ &  $\phantom{0}\phantom{0}\phantom{0}1.4\substack{+1.0\phantom{0}\phantom{0} \\ -0.3\phantom{0}\phantom{0}}$ &  $\phantom{0}\phantom{0}\phantom{0}0.6\substack{+0.5\phantom{0}\phantom{0} \\ -0.3\phantom{0}\phantom{0}}$ &  0.44 &  $26.7\substack{+15.3 \\ -32.4}$ \\[\dy]
4318\phantom{*} &  $0.70\substack{+0.15 \\ -0.05}$ &  $1.85\substack{+0.30 \\ -0.70}$ &  $\phantom{0}\phantom{0}11.9\substack{+1.9\phantom{0}\phantom{0} \\ -3.9\phantom{0}\phantom{0}}$ &  $\phantom{0}\phantom{0}\phantom{0}7.1\substack{+0.9\phantom{0}\phantom{0} \\ -5.7\phantom{0}\phantom{0}}$ &  0.60 &  $75.1\substack{+9.2\phantom{0} \\ -76.0}$ \\[\dy]
4320\phantom{*} &  $0.60\substack{+0.00 \\ -0.05}$ &  $0.30\substack{+0.15 \\ -0.00}$ &  $\phantom{0}\phantom{0}\phantom{0}3.1\substack{+2.6\phantom{0}\phantom{0} \\ -0.0\phantom{0}\phantom{0}}$ &  $\phantom{0}\phantom{0}\phantom{0}1.2\substack{+0.1\phantom{0}\phantom{0} \\ -0.7\phantom{0}\phantom{0}}$ &  0.39 &  $33.6\substack{+69.3 \\ -26.9}$ \\[\dy]
4321\phantom{*} &  $0.50\substack{+0.00 \\ -0.00}$ &  $0.80\substack{+0.05 \\ -0.00}$ &  $\phantom{0}\phantom{0}\phantom{0}5.9\substack{+0.6\phantom{0}\phantom{0} \\ -1.6\phantom{0}\phantom{0}}$ &  $\phantom{0}\phantom{0}\phantom{0}3.2\substack{+0.0\phantom{0}\phantom{0} \\ -2.5\phantom{0}\phantom{0}}$ &  0.55 &  $94.6\substack{+0.0\phantom{0} \\ -67.0}$ \\[\dy]
4322\phantom{*} &  $0.45\substack{+0.05 \\ -0.00}$ &  $0.75\substack{+0.05 \\ -0.15}$ &  $\phantom{0}\phantom{0}\phantom{0}4.2\substack{+2.9\phantom{0}\phantom{0} \\ -1.2\phantom{0}\phantom{0}}$ &  $\phantom{0}\phantom{0}\phantom{0}1.7\substack{+3.3\phantom{0}\phantom{0} \\ -1.1\phantom{0}\phantom{0}}$ &  0.40 &  $75.1\substack{+27.5 \\ -15.2}$ \\[\dy]
4330\phantom{*} &  $0.45\substack{+0.00 \\ -0.00}$ &  $0.75\substack{+0.05 \\ -0.05}$ &  $\phantom{0}\phantom{0}\phantom{0}5.7\substack{+2.5\phantom{0}\phantom{0} \\ -0.1\phantom{0}\phantom{0}}$ &  $\phantom{0}\phantom{0}\phantom{0}3.9\substack{+0.7\phantom{0}\phantom{0} \\ -2.7\phantom{0}\phantom{0}}$ &  0.68 &  $53.2\substack{+41.4 \\ -7.3\phantom{0}}$ \\[\dy]
4331\tablenotemark{*} &  $0.35\substack{+0.15 \\ -0.05}$ &  $1.90\substack{+0.15 \\ -0.15}$ &  $\phantom{0}\phantom{0}\phantom{0}2.5\substack{+1.8\phantom{0}\phantom{0} \\ -0.5\phantom{0}\phantom{0}}$ &  $\phantom{0}\phantom{0}\phantom{0}2.5\substack{+1.1\phantom{0}\phantom{0} \\ -2.0\phantom{0}\phantom{0}}$ &  1.00 &  $47.4\substack{+7.2\phantom{0} \\ -42.9}$ \\[\dy]
4333\tablenotemark{*} &  $0.40\substack{+0.05 \\ -0.05}$ &  $0.00\substack{+0.10 \\ -0.00}$ &  $\phantom{0}\phantom{0}\phantom{0}1.1\substack{+0.5\phantom{0}\phantom{0} \\ -0.0\phantom{0}\phantom{0}}$ &  $\phantom{0}\phantom{0}\phantom{0}1.0\substack{+0.1\phantom{0}\phantom{0} \\ -0.8\phantom{0}\phantom{0}}$ &  0.90 &  $16.8\substack{+12.2 \\ -8.5\phantom{0}}$ \\[\dy]
4335\phantom{*} &  $0.50\substack{+0.05 \\ -0.00}$ &  $0.05\substack{+0.05 \\ -0.00}$ &  $\phantom{0}\phantom{0}\phantom{0}3.1\substack{+3.5\phantom{0}\phantom{0} \\ -0.0\phantom{0}\phantom{0}}$ &  $\phantom{0}\phantom{0}\phantom{0}1.2\substack{+1.7\phantom{0}\phantom{0} \\ -0.6\phantom{0}\phantom{0}}$ &  0.38 &  $23.8\substack{+87.5 \\ -18.9}$ \\[\dy]
4337\phantom{*} &  $0.80\substack{+0.05 \\ -0.00}$ &  $2.00\substack{+0.00 \\ -0.10}$ &  $\phantom{0}\phantom{0}25.2\substack{+7.6\phantom{0}\phantom{0} \\ -1.2\phantom{0}\phantom{0}}$ &  $\phantom{0}\phantom{0}10.8\substack{+0.0\phantom{0}\phantom{0} \\ -5.7\phantom{0}\phantom{0}}$ &  0.43 &  $29.9\substack{+12.3 \\ -14.1}$ \\[\dy]
4339\phantom{*} &  $0.25\substack{+0.00 \\ -0.00}$ &  $1.05\substack{+0.00 \\ -0.05}$ &  $\phantom{0}\phantom{0}\phantom{0}8.8\substack{+3.1\phantom{0}\phantom{0} \\ -0.2\phantom{0}\phantom{0}}$ &  $\phantom{0}\phantom{0}\phantom{0}3.5\substack{+1.0\phantom{0}\phantom{0} \\ -1.7\phantom{0}\phantom{0}}$ &  0.39 &  $\phantom{0}8.4\substack{+6.6\phantom{0} \\ -6.2\phantom{0}}$ \\[\dy]
4345\tablenotemark{*} &  $0.25\substack{+0.05 \\ -0.00}$ &  $1.50\substack{+0.10 \\ -0.10}$ &  $\phantom{0}\phantom{0}\phantom{0}3.3\substack{+1.1\phantom{0}\phantom{0} \\ -0.2\phantom{0}\phantom{0}}$ &  $\phantom{0}\phantom{0}\phantom{0}3.2\substack{+0.0\phantom{0}\phantom{0} \\ -3.0\phantom{0}\phantom{0}}$ &  0.96 &  $16.8\substack{+17.0 \\ -15.4}$ \\[\dy]
4346\phantom{*} &  $0.45\substack{+0.00 \\ -0.00}$ &  $1.25\substack{+0.00 \\ -0.05}$ &  $\phantom{0}\phantom{0}41.8\substack{+0.9\phantom{0}\phantom{0} \\ -5.1\phantom{0}\phantom{0}}$ &  $\phantom{0}\phantom{0}12.6\substack{+0.0\phantom{0}\phantom{0} \\ -6.2\phantom{0}\phantom{0}}$ &  0.30 &  $42.2\substack{+18.5 \\ -14.8}$ \\[\dy]
4348\phantom{*} &  $0.50\substack{+0.00 \\ -0.00}$ &  $0.50\substack{+0.05 \\ -0.05}$ &  $\phantom{0}\phantom{0}\phantom{0}3.5\substack{+0.9\phantom{0}\phantom{0} \\ -0.4\phantom{0}\phantom{0}}$ &  $\phantom{0}\phantom{0}\phantom{0}2.2\substack{+0.0\phantom{0}\phantom{0} \\ -1.6\phantom{0}\phantom{0}}$ &  0.64 &  $59.7\substack{+1.0\phantom{0} \\ -51.3}$ \\[\dy]
4349\phantom{*} &  $0.70\substack{+0.05 \\ -0.00}$ &  $1.20\substack{+0.05 \\ -0.10}$ &  $\phantom{0}\phantom{0}\phantom{0}4.4\substack{+2.8\phantom{0}\phantom{0} \\ -0.0\phantom{0}\phantom{0}}$ &  $\phantom{0}\phantom{0}\phantom{0}2.4\substack{+0.9\phantom{0}\phantom{0} \\ -1.4\phantom{0}\phantom{0}}$ &  0.55 &  $\phantom{0}4.7\substack{+22.3 \\ -3.5\phantom{0}}$ \\[\dy]
4350\phantom{*} &  $0.60\substack{+0.05 \\ -0.00}$ &  $1.10\substack{+0.05 \\ -0.05}$ &  $\phantom{0}\phantom{0}\phantom{0}3.7\substack{+1.5\phantom{0}\phantom{0} \\ -0.0\phantom{0}\phantom{0}}$ &  $\phantom{0}\phantom{0}\phantom{0}1.6\substack{+1.4\phantom{0}\phantom{0} \\ -0.9\phantom{0}\phantom{0}}$ &  0.42 &  $11.9\substack{+22.2 \\ -10.3}$ \\[\dy]
4353\phantom{*} &  $0.75\substack{+0.15 \\ -0.05}$ &  $1.35\substack{+0.15 \\ -0.35}$ &  $\phantom{0}\phantom{0}\phantom{0}2.4\substack{+1.1\phantom{0}\phantom{0} \\ -0.2\phantom{0}\phantom{0}}$ &  $\phantom{0}\phantom{0}\phantom{0}1.0\substack{+0.5\phantom{0}\phantom{0} \\ -0.7\phantom{0}\phantom{0}}$ &  0.42 &  $37.6\substack{+1.5\phantom{0} \\ -49.6}$ \\[\dy]
4354\phantom{*} &  $0.45\substack{+0.00 \\ -0.00}$ &  $0.00\substack{+0.00 \\ -0.00}$ &  $\phantom{0}\phantom{0}\phantom{0}0.8\substack{+0.8\phantom{0}\phantom{0} \\ -0.0\phantom{0}\phantom{0}}$ &  $\phantom{0}\phantom{0}\phantom{0}0.4\substack{+0.0\phantom{0}\phantom{0} \\ -0.2\phantom{0}\phantom{0}}$ &  0.52 &  $\phantom{0}6.0\substack{+24.0 \\ -0.9\phantom{0}}$ \\[\dy]
4355\tablenotemark{*} &  $0.65\substack{+0.05 \\ -0.05}$ &  $1.05\substack{+0.15 \\ -0.05}$ &  $\phantom{0}\phantom{0}\phantom{0}2.0\substack{+0.9\phantom{0}\phantom{0} \\ -0.1\phantom{0}\phantom{0}}$ &  $\phantom{0}\phantom{0}\phantom{0}2.0\substack{+0.1\phantom{0}\phantom{0} \\ -1.8\phantom{0}\phantom{0}}$ &  1.00 &  $\phantom{0}9.5\substack{+2.0\phantom{0} \\ -6.6\phantom{0}}$ \\[\dy]
4360\tablenotemark{*} &  $0.35\substack{+0.05 \\ -0.05}$ &  $0.75\substack{+0.05 \\ -0.15}$ &  $\phantom{0}\phantom{0}\phantom{0}1.0\substack{+0.8\phantom{0}\phantom{0} \\ -0.1\phantom{0}\phantom{0}}$ &  $\phantom{0}\phantom{0}\phantom{0}1.0\substack{+0.0\phantom{0}\phantom{0} \\ -0.8\phantom{0}\phantom{0}}$ &  1.00 &  $\phantom{0}5.3\substack{+6.4\phantom{0} \\ -4.4\phantom{0}}$ \\[\dy]
4362\tablenotemark{*} &  $0.55\substack{+0.05 \\ -0.00}$ &  $0.65\substack{+0.05 \\ -0.05}$ &  $\phantom{0}\phantom{0}\phantom{0}1.2\substack{+0.7\phantom{0}\phantom{0} \\ -0.0\phantom{0}\phantom{0}}$ &  $\phantom{0}\phantom{0}\phantom{0}1.2\substack{+0.0\phantom{0}\phantom{0} \\ -1.0\phantom{0}\phantom{0}}$ &  1.00 &  $15.0\substack{+2.6\phantom{0} \\ -12.1}$ \\[\dy]
4364\phantom{*} &  $0.45\substack{+0.00 \\ -0.00}$ &  $0.05\substack{+0.00 \\ -0.05}$ &  $\phantom{0}\phantom{0}\phantom{0}0.8\substack{+0.6\phantom{0}\phantom{0} \\ -0.0\phantom{0}\phantom{0}}$ &  $\phantom{0}\phantom{0}\phantom{0}0.4\substack{+0.0\phantom{0}\phantom{0} \\ -0.2\phantom{0}\phantom{0}}$ &  0.53 &  $\phantom{0}7.5\substack{+55.9 \\ -2.2\phantom{0}}$ \\[\dy]
comb.\tablenotemark{e} &  $0.35\substack{+0.00 \\ -0.00}$ &  $1.45\substack{+0.00 \\ -0.00}$ &  $\phantom{0}188.9\substack{+12.6\phantom{0} \\ -15.1\phantom{0}}$ &  $\phantom{0}\phantom{0}61.6\substack{+0.0\phantom{0}\phantom{0} \\ -29.5\phantom{0}}$ &  0.33 &  $42.2\substack{+32.8 \\ -6.2\phantom{0}}$
\enddata
\tablenotetext{a}{Best-fit foreground and differential extinction parameters.
    Uncertainties are zero if the best-fit value equals the minimum or maximum
    estimate, or if there are no other solutions within $1\sigma$ of the
    best-fit SFH.
}
%\tablenotetext{b}{Mean total extinction in FUV, derived from $A_{\mathrm{V},f}
%    + dA_\mathrm{V}/2$ and the \citet{Cardelli89} extinction curve.
%}
\tablenotetext{b}{Total mass formed over the past $100\,\mathrm{Myr}$ of the
    SFHs. The corresponding mean SFR is $\langle \mathrm{SFR}\rangle_{100} =
    M_{100} \times 10^{-8}\,\mathrm{yr}^{-1}$.
}
\tablenotetext{c}{The mass of the age bin with the highest SFR over the last
    $100\,\mathrm{Myr}$ of the SFH at full time resolution.
}
\tablenotetext{d}{The mean age of the bin corresponding to $M_\mathrm{peak}$.}
\tablenotetext{e}{The combination of all regions except for region 4308.}
\tablenotetext{*}{SSP-like region (\S \ref{discussion.ssp}).}
\end{deluxetable*}

%% file: table3.tex
%%% For asymmetric uncertainties, `\substack{+x \\ -y}` from amsmath is
%%% slightly more compact than `^{+x}_{-y}`.

\begin{deluxetable*}{cccccc}
\tablecaption{FUV and NUV magnitudes modeled from the SFHs.\label{tab:fluxmod}}
\tablewidth{0pt}
\tablehead{
    \colhead{ID} &
    \colhead{$\mathrm{FUV_{SFH,0}}$\tablenotemark{a}} &
    \colhead{$\mathrm{(FUV-NUV)_{SFH,0}}$\tablenotemark{a}} &
    \colhead{$\mathrm{FUV_{SFH}}$\tablenotemark{b}} &
    \colhead{$\mathrm{(FUV-NUV)_{SFH}}$\tablenotemark{b}} &
    \colhead{$A_\mathrm{FUV}$\tablenotemark{c}} \\
    \colhead{} &
    \colhead{(AB mag)} &
    \colhead{(AB mag)} &
    \colhead{(AB mag)} &
    \colhead{(AB mag)} &
    \colhead{}
}
\startdata
4285\phantom{*} &  $17.51\substack{+0.02 \\ -0.23}$ &  $\phantom{+}0.03\substack{+0.01 \\ -0.00}$ &  $20.45\substack{+0.05 \\ -0.04}$ &  $\phantom{+}0.07\substack{+0.01 \\ -0.16}$ &  $2.94\substack{+0.00 \\ -0.24}$ \\[\dy]
4288\phantom{*} &  $17.18\substack{+0.44 \\ -0.11}$ &  $-0.14\substack{+0.04 \\ -0.00}$ &  $18.92\substack{+0.45 \\ -0.11}$ &  $-0.13\substack{+0.05 \\ -0.00}$ &  $1.73\substack{+0.00 \\ -0.00}$ \\[\dy]
4290\phantom{*} &  $14.63\substack{+0.06 \\ -0.40}$ &  $-0.15\substack{+0.04 \\ -0.01}$ &  $17.59\substack{+0.07 \\ -0.37}$ &  $-0.11\substack{+0.04 \\ -0.02}$ &  $2.96\substack{+0.03 \\ -0.12}$ \\[\dy]
4292\phantom{*} &  $16.59\substack{+0.95 \\ -0.32}$ &  $-0.00\substack{+0.02 \\ -0.11}$ &  $19.83\substack{+0.94 \\ -0.13}$ &  $\phantom{+}0.07\substack{+0.01 \\ -0.12}$ &  $3.24\substack{+0.04 \\ -0.19}$ \\[\dy]
4293\phantom{*} &  $15.43\substack{+0.55 \\ -0.05}$ &  $-0.05\substack{+0.00 \\ -0.06}$ &  $17.73\substack{+0.50 \\ -0.00}$ &  $-0.02\substack{+0.00 \\ -0.06}$ &  $2.30\substack{+0.09 \\ -0.04}$ \\[\dy]
4298\phantom{*} &  $15.11\substack{+0.51 \\ -0.02}$ &  $-0.09\substack{+0.00 \\ -0.07}$ &  $18.21\substack{+0.56 \\ -0.00}$ &  $-0.05\substack{+0.00 \\ -0.07}$ &  $3.10\substack{+0.03 \\ -0.14}$ \\[\dy]
4299\phantom{*} &  $17.11\substack{+0.90 \\ -0.00}$ &  $\phantom{+}0.11\substack{+0.00 \\ -0.16}$ &  $18.98\substack{+0.89 \\ -0.00}$ &  $\phantom{+}0.13\substack{+0.00 \\ -0.16}$ &  $1.87\substack{+0.05 \\ -0.00}$ \\[\dy]
4308\phantom{*} &  $10.81\substack{+0.08 \\ -0.03}$ &  $-0.08\substack{+0.01 \\ -0.01}$ &  $13.81\substack{+0.07 \\ -0.04}$ &  $-0.03\substack{+0.02 \\ -0.00}$ &  $3.00\substack{+0.00 \\ -0.00}$ \\[\dy]
4310\phantom{*} &  $17.21\substack{+0.92 \\ -0.12}$ &  $-0.01\substack{+0.01 \\ -0.10}$ &  $18.95\substack{+0.90 \\ -0.06}$ &  $\phantom{+}0.00\substack{+0.01 \\ -0.10}$ &  $1.74\substack{+0.09 \\ -0.06}$ \\[\dy]
4313\phantom{*} &  $17.29\substack{+0.77 \\ -1.15}$ &  $-0.14\substack{+0.09 \\ -0.01}$ &  $19.02\substack{+0.62 \\ -0.86}$ &  $-0.13\substack{+0.09 \\ -0.01}$ &  $1.73\substack{+0.21 \\ -0.29}$ \\[\dy]
4314\phantom{*} &  $14.59\substack{+0.04 \\ -0.76}$ &  $-0.26\substack{+0.12 \\ -0.00}$ &  $18.86\substack{+0.04 \\ -0.76}$ &  $-0.17\substack{+0.13 \\ -0.01}$ &  $4.27\substack{+0.11 \\ -0.11}$ \\[\dy]
4317\phantom{*} &  $16.89\substack{+0.28 \\ -0.77}$ &  $-0.20\substack{+0.14 \\ -0.00}$ &  $19.05\substack{+0.25 \\ -0.89}$ &  $-0.18\substack{+0.15 \\ -0.00}$ &  $2.16\substack{+0.27 \\ -0.09}$ \\[\dy]
4318\phantom{*} &  $16.75\substack{+0.30 \\ -0.16}$ &  $-0.01\substack{+0.01 \\ -0.10}$ &  $20.51\substack{+0.78 \\ -0.03}$ &  $\phantom{+}0.07\substack{+0.00 \\ -0.10}$ &  $3.76\substack{+0.17 \\ -0.15}$ \\[\dy]
4320\phantom{*} &  $16.52\substack{+0.52 \\ -0.00}$ &  $-0.11\substack{+0.02 \\ -0.02}$ &  $18.65\substack{+0.50 \\ -0.00}$ &  $-0.09\substack{+0.02 \\ -0.02}$ &  $2.14\substack{+0.06 \\ -0.02}$ \\[\dy]
4321\phantom{*} &  $16.94\substack{+0.81 \\ -0.00}$ &  $-0.03\substack{+0.00 \\ -0.09}$ &  $19.34\substack{+0.81 \\ -0.03}$ &  $\phantom{+}0.00\substack{+0.00 \\ -0.09}$ &  $2.40\substack{+0.05 \\ -0.00}$ \\[\dy]
4322\phantom{*} &  $17.45\substack{+1.03 \\ -0.06}$ &  $\phantom{+}0.01\substack{+0.01 \\ -0.12}$ &  $19.66\substack{+1.02 \\ -0.06}$ &  $\phantom{+}0.04\substack{+0.01 \\ -0.13}$ &  $2.21\substack{+0.14 \\ -0.10}$ \\[\dy]
4330\phantom{*} &  $16.87\substack{+0.82 \\ -0.05}$ &  $-0.02\substack{+0.00 \\ -0.08}$ &  $19.08\substack{+0.82 \\ -0.00}$ &  $\phantom{+}0.01\substack{+0.00 \\ -0.08}$ &  $2.21\substack{+0.05 \\ -0.05}$ \\[\dy]
4331\tablenotemark{*} &  $18.44\substack{+0.63 \\ -0.13}$ &  $\phantom{+}0.08\substack{+0.01 \\ -0.07}$ &  $21.23\substack{+0.26 \\ -0.01}$ &  $\phantom{+}0.12\substack{+0.01 \\ -0.06}$ &  $2.78\substack{+0.38 \\ -0.14}$ \\[\dy]
4333\tablenotemark{*} &  $18.18\substack{+0.81 \\ -0.27}$ &  $-0.07\substack{+0.02 \\ -0.07}$ &  $19.33\substack{+0.75 \\ -0.20}$ &  $-0.07\substack{+0.02 \\ -0.07}$ &  $1.16\substack{+0.15 \\ -0.07}$ \\[\dy]
4335\phantom{*} &  $17.03\substack{+0.84 \\ -0.00}$ &  $-0.06\substack{+0.00 \\ -0.07}$ &  $18.54\substack{+0.74 \\ -0.00}$ &  $-0.05\substack{+0.00 \\ -0.07}$ &  $1.51\substack{+0.14 \\ -0.00}$ \\[\dy]
4337\phantom{*} &  $15.14\substack{+0.09 \\ -0.07}$ &  $-0.04\substack{+0.00 \\ -0.00}$ &  $19.27\substack{+0.65 \\ -0.03}$ &  $\phantom{+}0.06\substack{+0.01 \\ -0.08}$ &  $4.13\substack{+0.12 \\ -0.06}$ \\[\dy]
4339\phantom{*} &  $14.79\substack{+0.15 \\ -0.32}$ &  $-0.15\substack{+0.02 \\ -0.02}$ &  $16.70\substack{+0.15 \\ -0.30}$ &  $-0.14\substack{+0.02 \\ -0.02}$ &  $1.91\substack{+0.00 \\ -0.04}$ \\[\dy]
4345\tablenotemark{*} &  $16.95\substack{+0.84 \\ -0.10}$ &  $-0.08\substack{+0.00 \\ -0.07}$ &  $19.20\substack{+0.78 \\ -0.03}$ &  $-0.06\substack{+0.00 \\ -0.07}$ &  $2.25\substack{+0.18 \\ -0.07}$ \\[\dy]
4346\phantom{*} &  $14.54\substack{+0.40 \\ -0.12}$ &  $-0.09\substack{+0.02 \\ -0.04}$ &  $17.18\substack{+0.40 \\ -0.08}$ &  $-0.05\substack{+0.02 \\ -0.04}$ &  $2.65\substack{+0.00 \\ -0.04}$ \\[\dy]
4348\phantom{*} &  $17.23\substack{+0.79 \\ -0.05}$ &  $-0.06\substack{+0.01 \\ -0.07}$ &  $19.32\substack{+0.78 \\ -0.01}$ &  $-0.04\substack{+0.01 \\ -0.07}$ &  $2.08\substack{+0.06 \\ -0.06}$ \\[\dy]
4349\phantom{*} &  $15.11\substack{+0.35 \\ -0.57}$ &  $-0.15\substack{+0.02 \\ -0.11}$ &  $18.43\substack{+0.20 \\ -0.57}$ &  $-0.09\substack{+0.03 \\ -0.11}$ &  $3.33\substack{+0.15 \\ -0.04}$ \\[\dy]
4350\phantom{*} &  $16.40\substack{+0.63 \\ -0.19}$ &  $-0.10\substack{+0.02 \\ -0.06}$ &  $19.35\substack{+0.57 \\ -0.14}$ &  $-0.05\substack{+0.02 \\ -0.06}$ &  $2.96\substack{+0.15 \\ -0.04}$ \\[\dy]
4353\phantom{*} &  $16.41\substack{+0.49 \\ -0.89}$ &  $-0.25\substack{+0.17 \\ -0.01}$ &  $20.00\substack{+0.27 \\ -0.87}$ &  $-0.19\substack{+0.20 \\ -0.00}$ &  $3.58\substack{+0.32 \\ -0.22}$ \\[\dy]
4354\phantom{*} &  $17.38\substack{+0.53 \\ -0.10}$ &  $-0.16\substack{+0.03 \\ -0.01}$ &  $18.68\substack{+0.53 \\ -0.10}$ &  $-0.15\substack{+0.03 \\ -0.01}$ &  $1.30\substack{+0.00 \\ -0.00}$ \\[\dy]
4355\tablenotemark{*} &  $16.73\substack{+0.95 \\ -0.08}$ &  $-0.13\substack{+0.00 \\ -0.07}$ &  $19.79\substack{+0.81 \\ -0.05}$ &  $-0.08\substack{+0.00 \\ -0.06}$ &  $3.06\substack{+0.19 \\ -0.10}$ \\[\dy]
4360\tablenotemark{*} &  $16.66\substack{+0.55 \\ -0.99}$ &  $-0.16\substack{+0.03 \\ -0.11}$ &  $18.58\substack{+0.46 \\ -0.83}$ &  $-0.15\substack{+0.04 \\ -0.11}$ &  $1.92\substack{+0.10 \\ -0.19}$ \\[\dy]
4362\tablenotemark{*} &  $17.87\substack{+1.23 \\ -0.00}$ &  $-0.08\substack{+0.00 \\ -0.09}$ &  $20.27\substack{+1.12 \\ -0.03}$ &  $-0.05\substack{+0.00 \\ -0.09}$ &  $2.39\substack{+0.14 \\ -0.00}$ \\[\dy]
4364\phantom{*} &  $18.03\substack{+0.69 \\ -0.14}$ &  $-0.15\substack{+0.03 \\ -0.01}$ &  $19.40\substack{+0.69 \\ -0.07}$ &  $-0.14\substack{+0.03 \\ -0.01}$ &  $1.37\substack{+0.00 \\ -0.07}$ \\[\dy]
comb.\tablenotemark{d} &  $12.03\substack{+0.10 \\ -0.11}$ &  $-0.13\substack{+0.01 \\ -0.02}$ &  $14.53\substack{+0.09 \\ -0.11}$ &  $-0.10\substack{+0.01 \\ -0.02}$ &  $2.50\substack{+0.00 \\ -0.00}$
\enddata
\tablenotetext{a}{Intrinsic (unreddened) FUV and NUV magnitudes modeled from
    the SFHs.
}
\tablenotetext{b}{Reddened FUV and NUV magnitudes modeled from the SFHs and the
    best-fit extinction parameters in Table \ref{tab:SFHdata}.
}
\tablenotetext{c}{FUV extinction correction, from the difference between
    $\mathrm{FUV_{SFH}}$ and $\mathrm{FUV_{SFH,0}}$. Uncertainties smaller than
    half the reported precision are rounded to zero.
}
\tablenotetext{d}{The combination of all regions except for region 4308.}
\tablenotetext{*}{SSP-like region (\S \ref{discussion.ssp}).}
\end{deluxetable*}

%% file: table4.tex
%%% For asymmetric uncertainties, `\substack{+x \\ -y}` from amsmath is
%%% slightly more compact than `^{+x}_{-y}`.

\begin{deluxetable*}{cccccccc}
\tablecaption{SFRs, ages, and masses from FUV and NUV
    fluxes.\label{tab:fluxderived}
}
\tablewidth{0pt}
\tablehead{
    \colhead{ID} &
    \colhead{$\mathrm{SFR_{FUV,0}}\tablenotemark{a}$} &
    \colhead{$\mathrm{SFR_{FUV}}\tablenotemark{b}$} &
    \colhead{$\mathrm{Age_{SSP}}\tablenotemark{c}$} &
    \colhead{$M_\mathrm{SSP}\tablenotemark{c}$} \\
    \colhead{} &
    \colhead{$(\times 10^{-5}\,M_\odot\,\mathrm{yr}^{-1})$}  &
    \colhead{$(\times 10^{-5}\,M_\odot\,\mathrm{yr}^{-1})$}  &
    \colhead{(Myr)} &
    \colhead{$(\times 10^3\,M_\odot)$}
}
\startdata
4285\phantom{*} &  $\phantom{0}\phantom{0}2.4\substack{+0.0\phantom{0} \\ -0.4\phantom{0}}$ &  $\phantom{0}\phantom{0}13.3\substack{+0.2 \\ -2.6}$ &  $161.6\substack{+8.9\phantom{0} \\ -8.9\phantom{0}}$ &  $\phantom{0}160.00\substack{+0.00\phantom{0}\phantom{0} \\ -0.00\phantom{0}\phantom{0}}$ \\[\dy]
4288\phantom{*} &  $\phantom{0}\phantom{0}3.2\substack{+1.4\phantom{0} \\ -0.5\phantom{0}}$ &  $\phantom{0}\phantom{0}\phantom{0}2.3\substack{+0.1 \\ -0.1}$ &  $\phantom{0}83.3\substack{+10.2 \\ -9.3\phantom{0}}$ &  $\phantom{0}\phantom{0}14.00\substack{+3.00\phantom{0}\phantom{0} \\ -2.00\phantom{0}\phantom{0}}$ \\[\dy]
4290\phantom{*} &  $\phantom{0}33.5\substack{+1.1\phantom{0} \\ -11.5}$ &  $\phantom{0}\phantom{0}40.7\substack{+1.1 \\ -4.2}$ &  $203.9\substack{+4.0\phantom{0} \\ -4.0\phantom{0}}$ &  $2000.00\substack{+0.00\phantom{0}\phantom{0} \\ -0.00\phantom{0}\phantom{0}}$ \\[\dy]
4292\phantom{*} &  $\phantom{0}\phantom{0}5.5\substack{+7.7\phantom{0} \\ -1.4\phantom{0}}$ &  $\phantom{0}\phantom{0}\phantom{0}3.1\substack{+0.2 \\ -0.5}$ &  $\phantom{0}\phantom{0}3.1\substack{+2.6\phantom{0} \\ -1.5\phantom{0}}$ &  $\phantom{0}\phantom{0}\phantom{0}1.30\substack{+1.70\phantom{0}\phantom{0} \\ \phantom{.}\phantom{0}}$ \\[\dy]
4293\phantom{*} &  $\phantom{0}16.0\substack{+9.8\phantom{0} \\ -0.7\phantom{0}}$ &  $\phantom{0}\phantom{0}10.8\substack{+0.9 \\ -0.5}$ &  $169.5\substack{+8.4\phantom{0} \\ -8.4\phantom{0}}$ &  $\phantom{0}900.00\substack{+0.00\phantom{0}\phantom{0} \\ -0.00\phantom{0}\phantom{0}}$ \\[\dy]
4298\phantom{*} &  $\phantom{0}21.6\substack{+12.9 \\ -0.4\phantom{0}}$ &  $\phantom{0}\phantom{0}29.8\substack{+0.9 \\ -3.7}$ &  $122.7\substack{+7.0\phantom{0} \\ -7.0\phantom{0}}$ &  $\phantom{0}150.00\substack{+0.00\phantom{0}\phantom{0} \\ -0.00\phantom{0}\phantom{0}}$ \\[\dy]
4299\phantom{*} &  $\phantom{0}\phantom{0}3.4\substack{+4.4\phantom{0} \\ -0.0\phantom{0}}$ &  $\phantom{0}\phantom{0}\phantom{0}2.5\substack{+0.1 \\ -0.1}$ &  $\phantom{0}61.1\substack{+12.6 \\ -12.9}$ &  $\phantom{0}\phantom{0}12.00\substack{+3.00\phantom{0}\phantom{0} \\ -2.20\phantom{0}\phantom{0}}$ \\[\dy]
4308\phantom{*} &  $1129.2\substack{+80.5 \\ -37.5}$ &  $1045.1\substack{+1.9 \\ -2.1}$ &  $\phantom{0}79.0\substack{+0.8\phantom{0} \\ -0.7\phantom{0}}$ &  $9800.00\substack{+0.00\phantom{0}\phantom{0} \\ -0.00\phantom{0}\phantom{0}}$ \\[\dy]
4310\phantom{*} &  $\phantom{0}\phantom{0}3.1\substack{+4.1\phantom{0} \\ -0.3\phantom{0}}$ &  $\phantom{0}\phantom{0}\phantom{0}2.0\substack{+0.2 \\ -0.1}$ &  $150.5\substack{+14.2 \\ -14.3}$ &  $\phantom{0}\phantom{0}77.00\substack{+0.00\phantom{0}\phantom{0} \\ -51.00\phantom{0}}$ \\[\dy]
4313\phantom{*} &  $\phantom{0}\phantom{0}2.9\substack{+2.6\phantom{0} \\ -1.9\phantom{0}}$ &  $\phantom{0}\phantom{0}\phantom{0}2.9\substack{+0.6 \\ -0.7}$ &  $141.2\substack{+11.7 \\ -11.6}$ &  $\phantom{0}\phantom{0}48.00\substack{+102.00 \\ -0.00\phantom{0}\phantom{0}}$ \\[\dy]
4314\phantom{*} &  $\phantom{0}34.6\substack{+1.3\phantom{0} \\ -17.6}$ &  $\phantom{0}\phantom{0}15.8\substack{+1.7 \\ -1.6}$ &  $\phantom{0}27.5\substack{+13.0 \\ -8.5\phantom{0}}$ &  $\phantom{0}\phantom{0}64.00\substack{+21.00\phantom{0} \\ -26.00\phantom{0}}$ \\[\dy]
4317\phantom{*} &  $\phantom{0}\phantom{0}4.2\substack{+1.0\phantom{0} \\ -2.4\phantom{0}}$ &  $\phantom{0}\phantom{0}\phantom{0}2.6\substack{+0.7 \\ -0.2}$ &  $121.8\substack{+15.5 \\ -15.5}$ &  $\phantom{0}\phantom{0}40.00\substack{+0.00\phantom{0}\phantom{0} \\ -0.00\phantom{0}\phantom{0}}$ \\[\dy]
4318\phantom{*} &  $\phantom{0}\phantom{0}4.8\substack{+1.5\phantom{0} \\ -0.6\phantom{0}}$ &  $\phantom{0}\phantom{0}\phantom{0}7.8\substack{+1.4 \\ -1.1}$ &  $\phantom{0}18.4\substack{+11.2 \\ -6.7\phantom{0}}$ &  $\phantom{0}\phantom{0}\phantom{0}1.60\substack{+1.20\phantom{0}\phantom{0} \\ -0.91\phantom{0}\phantom{0}}$ \\[\dy]
4320\phantom{*} &  $\phantom{0}\phantom{0}5.9\substack{+3.2\phantom{0} \\ -0.3\phantom{0}}$ &  $\phantom{0}\phantom{0}\phantom{0}8.9\substack{+0.5 \\ -0.2}$ &  $\phantom{0}84.7\substack{+5.4\phantom{0} \\ -5.4\phantom{0}}$ &  $\phantom{0}180.00\substack{+20.00\phantom{0} \\ -0.00\phantom{0}\phantom{0}}$ \\[\dy]
4321\phantom{*} &  $\phantom{0}\phantom{0}4.0\substack{+4.4\phantom{0} \\ -0.0\phantom{0}}$ &  $\phantom{0}\phantom{0}\phantom{0}2.6\substack{+0.1 \\ -0.1}$ &  $\phantom{0}66.2\substack{+14.2 \\ -16.8}$ &  $\phantom{0}\phantom{0}19.00\substack{+3.00\phantom{0}\phantom{0} \\ -7.00\phantom{0}\phantom{0}}$ \\[\dy]
4322\phantom{*} &  $\phantom{0}\phantom{0}2.5\substack{+3.8\phantom{0} \\ -0.1\phantom{0}}$ &  $\phantom{0}\phantom{0}\phantom{0}0.9\substack{+0.1 \\ -0.1}$ &  $\phantom{0}\phantom{0}5.6\substack{+7.8\phantom{0} \\ -3.5\phantom{0}}$ &  $\phantom{0}\phantom{0}\phantom{0}0.24\substack{+0.32\phantom{0}\phantom{0} \\ -0.12\phantom{0}\phantom{0}}$ \\[\dy]
4330\phantom{*} &  $\phantom{0}\phantom{0}4.2\substack{+4.8\phantom{0} \\ -0.2\phantom{0}}$ &  $\phantom{0}\phantom{0}\phantom{0}2.1\substack{+0.1 \\ -0.1}$ &  $\phantom{0}43.4\substack{+11.7 \\ -14.7}$ &  $\phantom{0}\phantom{0}23.00\substack{+16.00\phantom{0} \\ -6.00\phantom{0}\phantom{0}}$ \\[\dy]
4331\tablenotemark{*} &  $\phantom{0}\phantom{0}1.0\substack{+0.8\phantom{0} \\ -0.1\phantom{0}}$ &  $\phantom{0}\phantom{0}13.3\substack{+5.5 \\ -1.7}$ &  $130.9\substack{+8.0\phantom{0} \\ -8.1\phantom{0}}$ &  $2000.00\substack{+0.00\phantom{0}\phantom{0} \\ -0.00\phantom{0}\phantom{0}}$ \\[\dy]
4333\tablenotemark{*} &  $\phantom{0}\phantom{0}1.3\substack{+1.3\phantom{0} \\ -0.3\phantom{0}}$ &  $\phantom{0}\phantom{0}\phantom{0}2.1\substack{+0.3 \\ -0.1}$ &  $\phantom{0}\phantom{0}1.9\substack{+0.5\phantom{0} \\ -0.4\phantom{0}}$ &  $\phantom{0}\phantom{0}\phantom{0}0.40\substack{+0.00\phantom{0}\phantom{0} \\ -0.00\phantom{0}\phantom{0}}$ \\[\dy]
4335\phantom{*} &  $\phantom{0}\phantom{0}3.7\substack{+3.8\phantom{0} \\ -0.0\phantom{0}}$ &  $\phantom{0}\phantom{0}\phantom{0}5.2\substack{+0.7 \\ -0.1}$ &  $\phantom{0}36.8\substack{+5.4\phantom{0} \\ -7.4\phantom{0}}$ &  $\phantom{0}\phantom{0}46.00\substack{+0.00\phantom{0}\phantom{0} \\ -12.00\phantom{0}}$ \\[\dy]
4337\phantom{*} &  $\phantom{0}20.9\substack{+1.9\phantom{0} \\ -1.2\phantom{0}}$ &  $\phantom{0}\phantom{0}20.6\substack{+2.4 \\ -1.1}$ &  $\phantom{0}13.9\substack{+4.2\phantom{0} \\ -4.0\phantom{0}}$ &  $\phantom{0}\phantom{0}16.00\substack{+22.00\phantom{0} \\ -0.00\phantom{0}\phantom{0}}$ \\[\dy]
4339\phantom{*} &  $\phantom{0}28.9\substack{+3.0\phantom{0} \\ -8.5\phantom{0}}$ &  $\phantom{0}\phantom{0}31.9\substack{+0.2 \\ -1.3}$ &  $\phantom{0}63.5\substack{+3.1\phantom{0} \\ -3.1\phantom{0}}$ &  $\phantom{0}\phantom{0}96.00\substack{+24.00\phantom{0} \\ -0.00\phantom{0}\phantom{0}}$ \\[\dy]
4345\tablenotemark{*} &  $\phantom{0}\phantom{0}4.0\substack{+4.3\phantom{0} \\ -0.3\phantom{0}}$ &  $\phantom{0}\phantom{0}\phantom{0}6.7\substack{+1.2 \\ -0.4}$ &  $\phantom{0}\phantom{0}5.5\substack{+1.3\phantom{0} \\ -1.2\phantom{0}}$ &  $\phantom{0}\phantom{0}\phantom{0}0.46\substack{+0.28\phantom{0}\phantom{0} \\ -0.12\phantom{0}\phantom{0}}$ \\[\dy]
4346\phantom{*} &  $\phantom{0}36.5\substack{+16.4 \\ -3.8\phantom{0}}$ &  $\phantom{0}\phantom{0}36.1\substack{+0.3 \\ -1.3}$ &  $\phantom{0}18.3\substack{+1.6\phantom{0} \\ -1.7\phantom{0}}$ &  $\phantom{0}110.00\substack{+0.00\phantom{0}\phantom{0} \\ -0.00\phantom{0}\phantom{0}}$ \\[\dy]
4348\phantom{*} &  $\phantom{0}\phantom{0}3.1\substack{+3.2\phantom{0} \\ -0.1\phantom{0}}$ &  $\phantom{0}\phantom{0}\phantom{0}5.3\substack{+0.3 \\ -0.3}$ &  $\phantom{0}\phantom{0}9.3\substack{+2.9\phantom{0} \\ -1.9\phantom{0}}$ &  $\phantom{0}\phantom{0}\phantom{0}0.76\substack{+0.10\phantom{0}\phantom{0} \\ -0.25\phantom{0}\phantom{0}}$ \\[\dy]
4349\phantom{*} &  $\phantom{0}21.6\substack{+8.1\phantom{0} \\ -8.9\phantom{0}}$ &  $\phantom{0}\phantom{0}\phantom{0}9.6\substack{+1.4 \\ -0.4}$ &  $\phantom{0}80.6\substack{+9.6\phantom{0} \\ -9.2\phantom{0}}$ &  $\phantom{0}190.00\substack{+40.00\phantom{0} \\ -30.00\phantom{0}}$ \\[\dy]
4350\phantom{*} &  $\phantom{0}\phantom{0}6.6\substack{+4.9\phantom{0} \\ -1.0\phantom{0}}$ &  $\phantom{0}\phantom{0}\phantom{0}5.7\substack{+0.8 \\ -0.3}$ &  $\phantom{0}41.4\substack{+6.9\phantom{0} \\ -12.4}$ &  $\phantom{0}\phantom{0}23.00\substack{+9.00\phantom{0}\phantom{0} \\ -6.00\phantom{0}\phantom{0}}$ \\[\dy]
4353\phantom{*} &  $\phantom{0}\phantom{0}6.5\substack{+3.5\phantom{0} \\ -4.3\phantom{0}}$ &  $\phantom{0}\phantom{0}\phantom{0}5.5\substack{+1.9 \\ -1.0}$ &  $\phantom{0}\phantom{0}2.3\substack{+1.9\phantom{0} \\ -0.8\phantom{0}}$ &  $\phantom{0}\phantom{0}\phantom{0}0.21\substack{+0.05\phantom{0}\phantom{0} \\ \phantom{.}\phantom{0}}$ \\[\dy]
4354\phantom{*} &  $\phantom{0}\phantom{0}2.7\substack{+1.3\phantom{0} \\ -0.5\phantom{0}}$ &  $\phantom{0}\phantom{0}\phantom{0}3.3\substack{+0.0 \\ -0.0}$ &  $\phantom{0}45.9\substack{+3.8\phantom{0} \\ -3.8\phantom{0}}$ &  $\phantom{0}\phantom{0}17.00\substack{+0.00\phantom{0}\phantom{0} \\ -4.00\phantom{0}\phantom{0}}$ \\[\dy]
4355\tablenotemark{*} &  $\phantom{0}\phantom{0}4.8\substack{+5.8\phantom{0} \\ -0.4\phantom{0}}$ &  $\phantom{0}\phantom{0}10.0\substack{+1.9 \\ -0.9}$ &  $\phantom{0}\phantom{0}1.1\substack{+0.3\phantom{0} \\ \phantom{.}\phantom{0}}$ &  $\phantom{0}\phantom{0}\phantom{0}1.90\substack{+0.00\phantom{0}\phantom{0} \\ \phantom{.}\phantom{0}}$ \\[\dy]
4360\tablenotemark{*} &  $\phantom{0}\phantom{0}5.2\substack{+3.2\phantom{0} \\ -3.7\phantom{0}}$ &  $\phantom{0}\phantom{0}\phantom{0}5.4\substack{+0.5 \\ -0.9}$ &  $\phantom{0}\phantom{0}2.2\substack{+0.8\phantom{0} \\ -0.4\phantom{0}}$ &  $\phantom{0}\phantom{0}\phantom{0}0.28\substack{\phantom{.}\phantom{0} \\ -0.00\phantom{0}\phantom{0}}$ \\[\dy]
4362\tablenotemark{*} &  $\phantom{0}\phantom{0}1.7\substack{+2.9\phantom{0} \\ -0.0\phantom{0}}$ &  $\phantom{0}\phantom{0}\phantom{0}3.9\substack{+0.6 \\ -0.1}$ &  $\phantom{0}29.5\substack{+11.6 \\ -8.3\phantom{0}}$ &  $\phantom{0}\phantom{0}\phantom{0}5.80\substack{+2.00\phantom{0}\phantom{0} \\ -2.30\phantom{0}\phantom{0}}$ \\[\dy]
4364\phantom{*} &  $\phantom{0}\phantom{0}1.5\substack{+1.3\phantom{0} \\ -0.2\phantom{0}}$ &  $\phantom{0}\phantom{0}\phantom{0}1.5\substack{+0.0 \\ -0.1}$ &  $\phantom{0}76.1\substack{+9.6\phantom{0} \\ -9.8\phantom{0}}$ &  $\phantom{0}\phantom{0}\phantom{0}9.80\substack{+2.20\phantom{0}\phantom{0} \\ -1.50\phantom{0}\phantom{0}}$ \\[\dy]
comb.\tablenotemark{d} &  $368.8\substack{+31.8 \\ -36.4}$ &  $\phantom{0}297.3\substack{+0.9 \\ -0.9}$ &   &  
\enddata
\tablenotetext{a}{SFR derived using the modeled intrinsic FUV magnitudes and
    the flux calibration from \citet{Kennicutt12}.
}
\tablenotetext{b}{SFR derived using the extinction-corrected observed FUV
    magnitudes and the flux calibration from \citet{Kennicutt12}. Uncertainties
    smaller than half the reported precision are rounded to zero.
}\tablenotetext{c}{SSP ages and masses from \citet{Kang09}. Missing
    uncertainties indicate that the minimum/maximum value either is not
    available (4355) or is larger/smaller than the best value (4292, 4353,
    4360). The uncertainty is zero where the minimum/maximum value equals the
    best value.
}
\tablenotetext{d}{The combination of all regions except for region 4308.}
\tablenotetext{*}{SSP-like region (\S \ref{discussion.ssp}).}
\end{deluxetable*}

%% file: uv_regions_v6.5.bbl
\begin{thebibliography}{widest-label}

\bibitem[Astropy Collaboration et al.(2013)]{astropy} Astropy Collaboration, Robitaille, T.~P., Tollerud, E.~J., et al.\ 2013, \aap, 558, A33
\bibitem[Ballesteros-Paredes et al.(2011)]{Ballesteros11} Ballesteros-Paredes, J., V{\'a}zquez-Semadeni, E., Gazol, A., et al.\ 2011, \mnras, 416, 1436
\bibitem[Barmby et al.(2000)]{Barmby00} Barmby, P., Huchra, J.~P., Brodie, J.~P., et al.\ 2000, \aj, 119, 727
\bibitem[Barnes et al.(2011)]{Barnes11} Barnes, K.~L., van Zee, L., \& Skillman, E.~D.\ 2011, \apj, 743, 137
\bibitem[Berkhuijsen \& Fletcher(2008)]{Berkhuijsen08} Berkhuijsen, E.~M., \& Fletcher, A.\ 2008, \mnras, 390, L19
\bibitem[Bianchi et al.(1996)]{Bianchi96} Bianchi, L., Clayton, G.~C., Bohlin, R.~C., Hutchings, J.~B., \& Massey, P.\ 1996, \apj, 471, 203
\bibitem[Bianchi(2011)]{Bianchi11} Bianchi, L.\ 2011, \apss, 335, 51
\bibitem[Bressan et al.(2012)]{Bressan12} Bressan, A., Marigo, P., Girardi, L., et al.\ 2012, \mnras, 427, 127
\bibitem[Cardelli et al.(1989)]{Cardelli89} Cardelli, J.~A., Clayton, G.~C., \& Mathis, J.~S.\ 1989, \apj, 345, 245
\bibitem[Conroy et al.(2009)]{Conroy09} Conroy, C., Gunn, J.~E., \& White, M.\ 2009, \apj, 699, 486
\bibitem[Conroy \& Gunn(2010)]{Conroy10} Conroy, C., \& Gunn, J.~E.\ 2010, \apj, 712, 833
\bibitem[Dalcanton et al.(2012)]{Dalcanton12} Dalcanton, J.~J., Williams, B.~F., Lang, D., et al.\ 2012, \apjs, 200, 18
\bibitem[Dalcanton et al.(2014)]{Dalcanton14} Dalcanton, J.~J., Fouesneau, M., Hogg, D.~W., et al.\ 2014, preprint
\bibitem[Dolphin(2000)]{Dolphin00} Dolphin, A.~E.\ 2000, \pasp, 112, 1383
\bibitem[Dolphin(2002)]{Dolphin02} Dolphin, A.~E.\ 2002, \mnras, 332, 91
\bibitem[Dolphin(2013)]{Dolphin13} Dolphin, A.~E.\ 2013, \apj, 775, 76
\bibitem[Efremova et al.(2011)]{Efremova11} Efremova, B.~V., Bianchi, L., Thilker, D.~A., et al.\ 2011, \apj, 730, 88
\bibitem[Fitzpatrick \& Massa(2007)]{Fitzpatrick07} Fitzpatrick, E.~L., \& Massa, D.\ 2007, \apj, 663, 320
\bibitem[Girardi et al.(2005)]{Girardi05} Girardi, L., Groenewegen, M.~A.~T., Hatziminaoglou, E., \& da Costa, L.\ 2005, \aap, 436, 895
\bibitem[Girardi et al.(2010)]{Girardi10} Girardi, L., Williams, B.~F., Gilbert, K.~M., et al.\ 2010, \apj, 724, 1030
\bibitem[Gogarten et al.(2009)]{Gogarten09} Gogarten, S.~M., Dalcanton, J.~J., Williams, B.~F., et al.\ 2009, \apj, 691, 115
\bibitem[Hao et al.(2011)]{Hao11} Hao, C.-N., Kennicutt, R.~C., Johnson, B.~D., et al.\ 2011, \apj, 741, 124
\bibitem[Hill et al.(2008)]{Hill08} Hill, A.~S., Benjamin, R.~A., Kowal, G., et al.\ 2008, \apj, 686, 363
\bibitem[Hunter(2007)]{matplotlib} Hunter, J.~D.\ 2007, Computing in Science \& Engineering, 9, 3
\bibitem[Johnson et al.(2013)]{Johnson13} Johnson, B.~D., Weisz, D.~R., Dalcanton, J.~J., et al.\ 2013, \apj, 772, 8
\bibitem[Kang et al.(2009)]{Kang09} Kang, Y., Bianchi, L., \& Rey, S.-C.\ 2009, \apj, 703, 614
\bibitem[Kennicutt(1998)]{Kennicutt98} Kennicutt, R.~C., Jr.\ 1998, \araa, 36, 189
\bibitem[Kennicutt \& Evans(2012)]{Kennicutt12} Kennicutt, R.~C., \& Evans, N.~J.\ 2012, \araa, 50, 531
\bibitem[Kroupa(2001)]{Kroupa01} Kroupa, P.\ 2001, \mnras, 322, 231
\bibitem[Leitherer et al.(1999)]{Leitherer99} Leitherer, C., Schaerer, D., Goldader, J.~D., et al.\ 1999, \apjs, 123, 3
\bibitem[Leroy et al.(2012)]{Leroy12} Leroy, A.~K., Bigiel, F., de Blok, W.~J.~G., et al.\ 2012, \aj, 144, 3
\bibitem[McConnachie et al.(2005)]{McConnachie05} McConnachie, A.~W., Irwin, M.~J., Ferguson, A.~M.~N., et al.\ 2005, \mnras, 356, 979
\bibitem[Morrissey et al.(2007)]{Morrissey07} Morrissey, P., Conrow, T., Barlow, T.~A., et al.\ 2007, \apjs, 173, 682
\bibitem[Murphy et al.(2011)]{Murphy11} Murphy, E.~J., Condon, J.~J., Schinnerer, E., et al.\ 2011, \apj, 737, 67
\bibitem[Oliphant(2007)]{scipy} Oliphant, T.~E.\ 2007, Computing in Science \& Engineering, 9, 3
\bibitem[P\'erez \& Granger(2007)]{ipython} P\'erez, F., Granger, B.~E.\ 2007, Computing in Science and Engineering, 9, 3
\bibitem[Shetty et al.(2011)]{Shetty11} Shetty, R., Glover, S.~C., Dullemond, C.~P., \& Klessen, R.~S.\ 2011, \mnras, 412, 1686
\bibitem[Tully(1994)]{Tully94} Tully, R.~B.\ 1994, VizieR Online Data Catalog, 7145, 0

\end{thebibliography}
